%
%
%
%
%
%
%
%
\def\standardrisposta{s }\def\reducedrisposta{r }
\def\mplarisposta{mpla }\def\zerorisposta{z }
\def\doublerisposta{d }\def\cartarisposta{e }\def\amsrisposta{y }
\newcount\ingrandimento \newcount\sinnota \newcount\dimnota
\newcount\unoduecol \newdimen\collhsize \newdimen\tothsize
\newdimen\fullhsize \newcount\controllorisposta \sinnota=1
\newskip\infralinea  \global\controllorisposta=0
%
%
%
%
%
\def\risposta{s }
\def\srisposta{e }
\def\arisposta{y }
\ifx\risposta\standardrisposta \ingrandimento=1200
\message {>> This will come out UNREDUCED << }
\dimnota=2 \unoduecol=1 \global\controllorisposta=1 \fi
\ifx\risposta\reducedrisposta \ingrandimento=1095 \dimnota=1
\unoduecol=1  \global\controllorisposta=1
\message {>> This will come out REDUCED << } \fi
\ifx\risposta\doublerisposta \ingrandimento=1000 \dimnota=2
\unoduecol=2   \message {>> You must print this in
LANDSCAPE orientation << } \global\controllorisposta=1 \fi
\ifx\risposta\mplarisposta \ingrandimento=1000 \dimnota=1
\message {>> Mod. Phys. Lett. A format << }
\unoduecol=1 \global\controllorisposta=1 \fi
\ifx\risposta\zerorisposta \ingrandimento=1000 \dimnota=2
\message {>> Zero Magnification format << }
\unoduecol=1 \global\controllorisposta=1 \fi
\ifnum\controllorisposta=0  \ingrandimento=1200
\message {>>> ERROR IN INPUT, I ASSUME STANDARD
UNREDUCED FORMAT <<< }  \dimnota=2 \unoduecol=1 \fi
\magnification=\ingrandimento
%
%
%
%
\newdimen\eucolumnsize \newdimen\eudoublehsize \newdimen\eudoublevsize
\newdimen\uscolumnsize \newdimen\usdoublehsize \newdimen\usdoublevsize
\newdimen\eusinglehsize \newdimen\eusinglevsize \newdimen\ussinglehsize
\newskip\standardbaselineskip \newdimen\ussinglevsize
\newskip\reducedbaselineskip \newskip\doublebaselineskip
\eucolumnsize=12.0truecm    
\eudoublehsize=25.5truecm   
\eudoublevsize=6.5truein    
\uscolumnsize=4.4truein     
\usdoublehsize=9.4truein    
\usdoublevsize=6.8truein    
\eusinglehsize=6.5truein    
\eusinglevsize=24truecm     
\ussinglehsize=6.5truein    
\ussinglevsize=8.9truein    
\standardbaselineskip=16pt plus.2pt  
\reducedbaselineskip=14pt plus.2pt   
\doublebaselineskip=12pt plus.2pt    
%
%
\def\Portoffset{}
\def\Landoffset{}
\ifx\risposta\mplarisposta \def\Portoffset{\hoffset=1.8truecm} \fi
%
%
\def\Landspec{}
\tolerance=10000
\parskip=0pt plus2pt  \leftskip=0pt \rightskip=0pt
%
%
\ifx\risposta\standardrisposta \infralinea=\standardbaselineskip \fi
\ifx\risposta\reducedrisposta  \infralinea=\reducedbaselineskip \fi
\ifx\risposta\doublerisposta   \infralinea=\doublebaselineskip \fi
\ifx\risposta\mplarisposta     \infralinea=13pt \fi
\ifx\risposta\zerorisposta     \infralinea=12pt plus.2pt\fi
\ifnum\controllorisposta=0    \infralinea=\standardbaselineskip \fi
\ifx\risposta\doublerisposta   \Landoffset \else \Portoffset \fi
\ifx\risposta\doublerisposta \ifx\srisposta\cartarisposta
\tothsize=\eudoublehsize \collhsize=\eucolumnsize
\vsize=\eudoublevsize  \else  \tothsize=\usdoublehsize
\collhsize=\uscolumnsize \vsize=\usdoublevsize \fi \else
\ifx\srisposta\cartarisposta \tothsize=\eusinglehsize
\vsize=\eusinglevsize \else  \tothsize=\ussinglehsize
\vsize=\ussinglevsize \fi \collhsize=4.4truein \fi
\ifx\risposta\mplarisposta \tothsize=5.0truein
\vsize=7.8truein \collhsize=4.4truein \fi
%
%
%
%
\newcount\contaeuler \newcount\contacyrill \newcount\contaams
\font\ninerm=cmr9  \font\eightrm=cmr8  \font\sixrm=cmr6
\font\ninei=cmmi9  \font\eighti=cmmi8  \font\sixi=cmmi6
\font\ninesy=cmsy9  \font\eightsy=cmsy8  \font\sixsy=cmsy6
\font\ninebf=cmbx9  \font\eightbf=cmbx8  \font\sixbf=cmbx6
\font\ninett=cmtt9  \font\eighttt=cmtt8  \font\nineit=cmti9
\font\eightit=cmti8 \font\ninesl=cmsl9  \font\eightsl=cmsl8
\skewchar\ninei='177 \skewchar\eighti='177 \skewchar\sixi='177
\skewchar\ninesy='60 \skewchar\eightsy='60 \skewchar\sixsy='60
\hyphenchar\ninett=-1 \hyphenchar\eighttt=-1 \hyphenchar\tentt=-1
%
\font\tencmmib=cmmib10  \newfam\cmmibfam  \skewchar\tencmmib='177
\font\tencmbsy=cmbsy10  \newfam\cmbsyfam  \skewchar\tencmbsy='60
\def\scaps{\cmcsc}                 
\font\tencmcsc=cmcsc10  \newfam\cmcscfam
\ifnum\ingrandimento=1095

\font\capsone=cmcsc10 at 10.95pt 

\else

\font\capsone=cmcsc10 at 12pt 
\fi

\def\ttaarr{\bf}		
\def\ppaarr{\sl}		

%
%
%
\newfam\eufmfam \newfam\msamfam \newfam\msbmfam \newfam\eufbfam
\def\Loadeulerfonts{\global\contaeuler=1 \ifx\arisposta\amsrisposta
\font\teneufm=eufm10              
\font\eighteufm=eufm8 \font\nineeufm=eufm9 \font\sixeufm=eufm6
\font\seveneufm=eufm7  \font\fiveeufm=eufm5
\font\teneufb=eufb10              
\font\eighteufb=eufb8 \font\nineeufb=eufb9 \font\sixeufb=eufb6
\font\seveneufb=eufb7  \font\fiveeufb=eufb5
\font\teneurm=eurm10              
\font\eighteurm=eurm8 \font\nineeurm=eurm9
\font\teneurb=eurb10              
\font\eighteurb=eurb8 \font\nineeurb=eurb9
\font\teneusm=eusm10              
\font\eighteusm=eusm8 \font\nineeusm=eusm9
\font\teneusb=eusb10              
\font\eighteusb=eusb8 \font\nineeusb=eusb9
\else \def\eufm{\tt} \def\eufb{\tt} \def\eurm{\tt} \def\eurb{\tt}
\def\eusm{\tt} \def\eusb{\tt}    \fi}
\def\loadeuler{\Loadeulerfonts\tenpoint}
\def\loadamsmath{\global\contaams=1 \ifx\arisposta\amsrisposta
\font\tenmsam=msam10 \font\ninemsam=msam9 \font\eightmsam=msam8
\font\sevenmsam=msam7 \font\sixmsam=msam6 \font\fivemsam=msam5
\font\tenmsbm=msbm10 \font\ninemsbm=msbm9 \font\eightmsbm=msbm8
\font\sevenmsbm=msbm7 \font\sixmsbm=msbm6 \font\fivemsbm=msbm5
\else \def\msbm{\bf} \fi \def\Bbb{\msbm} \def\symbl{\msam} \tenpoint}
\def\loadcyrill{\global\contacyrill=1 \ifx\arisposta\amsrisposta
\font\tenwncyr=wncyr10 \font\ninewncyr=wncyr9 \font\eightwncyr=wncyr8
\font\tenwncyb=wncyr10 \font\ninewncyb=wncyr9 \font\eightwncyb=wncyr8
\font\tenwncyi=wncyr10 \font\ninewncyi=wncyr9 \font\eightwncyi=wncyr8
\else \def\cyrill{\sl} \def\cyrilb{\sl} \def\cyrili{\sl} \fi\tenpoint}
\ifx\arisposta\amsrisposta
\font\sevenex=cmex7               
\font\eightex=cmex8  \font\nineex=cmex9
\font\ninecmmib=cmmib9   \font\eightcmmib=cmmib8
\font\sevencmmib=cmmib7 \font\sixcmmib=cmmib6
\font\fivecmmib=cmmib5   \skewchar\ninecmmib='177
\skewchar\eightcmmib='177  \skewchar\sevencmmib='177
\skewchar\sixcmmib='177   \skewchar\fivecmmib='177
\font\ninecmbsy=cmbsy9    \font\eightcmbsy=cmbsy8
\font\sevencmbsy=cmbsy7  \font\sixcmbsy=cmbsy6
\font\fivecmbsy=cmbsy5   \skewchar\ninecmbsy='60
\skewchar\eightcmbsy='60  \skewchar\sevencmbsy='60
\skewchar\sixcmbsy='60    \skewchar\fivecmbsy='60
\font\ninecmcsc=cmcsc9    \font\eightcmcsc=cmcsc8     \else
\def\cmmib{\fam\cmmibfam\tencmmib}\textfont\cmmibfam=\tencmmib
\scriptfont\cmmibfam=\tencmmib \scriptscriptfont\cmmibfam=\tencmmib
\def\cmbsy{\fam\cmbsyfam\tencmbsy} \textfont\cmbsyfam=\tencmbsy
\scriptfont\cmbsyfam=\tencmbsy \scriptscriptfont\cmbsyfam=\tencmbsy
\scriptfont\cmcscfam=\tencmcsc \scriptscriptfont\cmcscfam=\tencmcsc
\def\cmcsc{\fam\cmcscfam\tencmcsc} \textfont\cmcscfam=\tencmcsc \fi
\catcode`@=11
\newskip\ttglue
\gdef\tenpoint{\def\rm{\fam0\tenrm}
  \textfont0=\tenrm \scriptfont0=\sevenrm \scriptscriptfont0=\fiverm
  \textfont1=\teni \scriptfont1=\seveni \scriptscriptfont1=\fivei
  \textfont2=\tensy \scriptfont2=\sevensy \scriptscriptfont2=\fivesy
  \textfont3=\tenex \scriptfont3=\tenex \scriptscriptfont3=\tenex
  \def\mcal{\fam2 \tensy}  \def\mmit{\fam1 \teni}
  \textfont\itfam=\tenit \def\it{\fam\itfam\tenit}
  \textfont\slfam=\tensl \def\sl{\fam\slfam\tensl}
  \textfont\ttfam=\tentt \scriptfont\ttfam=\eighttt
  \scriptscriptfont\ttfam=\eighttt  \def\tt{\fam\ttfam\tentt}
  \textfont\bffam=\tenbf \scriptfont\bffam=\sevenbf
  \scriptscriptfont\bffam=\fivebf \def\bf{\fam\bffam\tenbf}
     \ifx\arisposta\amsrisposta    \ifnum\contaeuler=1
  \textfont\eufmfam=\teneufm \scriptfont\eufmfam=\seveneufm
  \scriptscriptfont\eufmfam=\fiveeufm \def\eufm{\fam\eufmfam\teneufm}
  \textfont\eufbfam=\teneufb \scriptfont\eufbfam=\seveneufb
  \scriptscriptfont\eufbfam=\fiveeufb \def\eufb{\fam\eufbfam\teneufb}
  \def\eurm{\teneurm} \def\eurb{\teneurb} \def\eusm{\teneusm}
  \def\eusb{\teneusb}    \fi    \ifnum\contaams=1
  \textfont\msamfam=\tenmsam \scriptfont\msamfam=\sevenmsam
  \scriptscriptfont\msamfam=\fivemsam \def\msam{\fam\msamfam\tenmsam}
  \textfont\msbmfam=\tenmsbm \scriptfont\msbmfam=\sevenmsbm
  \scriptscriptfont\msbmfam=\fivemsbm \def\msbm{\fam\msbmfam\tenmsbm}
     \fi      \ifnum\contacyrill=1     \def\cyrill{\tenwncyr}
  \def\cyrilb{\tenwncyb}  \def\cyrili{\tenwncyi}         \fi
  \textfont3=\tenex \scriptfont3=\sevenex \scriptscriptfont3=\sevenex
  \def\cmmib{\fam\cmmibfam\tencmmib} \scriptfont\cmmibfam=\sevencmmib
  \textfont\cmmibfam=\tencmmib  \scriptscriptfont\cmmibfam=\fivecmmib
  \def\cmbsy{\fam\cmbsyfam\tencmbsy} \scriptfont\cmbsyfam=\sevencmbsy
  \textfont\cmbsyfam=\tencmbsy  \scriptscriptfont\cmbsyfam=\fivecmbsy
  \def\cmcsc{\fam\cmcscfam\tencmcsc} \scriptfont\cmcscfam=\eightcmcsc
  \textfont\cmcscfam=\tencmcsc \scriptscriptfont\cmcscfam=\eightcmcsc
     \fi            \tt \ttglue=.5em plus.25em minus.15em
  \normalbaselineskip=12pt
  \setbox\strutbox=\hbox{\vrule height8.5pt depth3.5pt width0pt}
  \let\sc=\eightrm \let\big=\tenbig   \normalbaselines
  \baselineskip=\infralinea  \rm}
\gdef\ninepoint{\def\rm{\fam0\ninerm}
  \textfont0=\ninerm \scriptfont0=\sixrm \scriptscriptfont0=\fiverm
  \textfont1=\ninei \scriptfont1=\sixi \scriptscriptfont1=\fivei
  \textfont2=\ninesy \scriptfont2=\sixsy \scriptscriptfont2=\fivesy
  \textfont3=\tenex \scriptfont3=\tenex \scriptscriptfont3=\tenex
  \def\mcal{\fam2 \ninesy}  \def\mmit{\fam1 \ninei}
  \textfont\itfam=\nineit \def\it{\fam\itfam\nineit}
  \textfont\slfam=\ninesl \def\sl{\fam\slfam\ninesl}
  \textfont\ttfam=\ninett \scriptfont\ttfam=\eighttt
  \scriptscriptfont\ttfam=\eighttt \def\tt{\fam\ttfam\ninett}
  \textfont\bffam=\ninebf \scriptfont\bffam=\sixbf
  \scriptscriptfont\bffam=\fivebf \def\bf{\fam\bffam\ninebf}
     \ifx\arisposta\amsrisposta  \ifnum\contaeuler=1
  \textfont\eufmfam=\nineeufm \scriptfont\eufmfam=\sixeufm
  \scriptscriptfont\eufmfam=\fiveeufm \def\eufm{\fam\eufmfam\nineeufm}
  \textfont\eufbfam=\nineeufb \scriptfont\eufbfam=\sixeufb
  \scriptscriptfont\eufbfam=\fiveeufb \def\eufb{\fam\eufbfam\nineeufb}
  \def\eurm{\nineeurm} \def\eurb{\nineeurb} \def\eusm{\nineeusm}
  \def\eusb{\nineeusb}     \fi   \ifnum\contaams=1
  \textfont\msamfam=\ninemsam \scriptfont\msamfam=\sixmsam
  \scriptscriptfont\msamfam=\fivemsam \def\msam{\fam\msamfam\ninemsam}
  \textfont\msbmfam=\ninemsbm \scriptfont\msbmfam=\sixmsbm
  \scriptscriptfont\msbmfam=\fivemsbm \def\msbm{\fam\msbmfam\ninemsbm}
     \fi       \ifnum\contacyrill=1     \def\cyrill{\ninewncyr}
  \def\cyrilb{\ninewncyb}  \def\cyrili{\ninewncyi}         \fi
  \textfont3=\nineex \scriptfont3=\sevenex \scriptscriptfont3=\sevenex
  \def\cmmib{\fam\cmmibfam\ninecmmib}  \textfont\cmmibfam=\ninecmmib
  \scriptfont\cmmibfam=\sixcmmib \scriptscriptfont\cmmibfam=\fivecmmib
  \def\cmbsy{\fam\cmbsyfam\ninecmbsy}  \textfont\cmbsyfam=\ninecmbsy
  \scriptfont\cmbsyfam=\sixcmbsy \scriptscriptfont\cmbsyfam=\fivecmbsy
  \def\cmcsc{\fam\cmcscfam\ninecmcsc} \scriptfont\cmcscfam=\eightcmcsc
  \textfont\cmcscfam=\ninecmcsc \scriptscriptfont\cmcscfam=\eightcmcsc
     \fi            \tt \ttglue=.5em plus.25em minus.15em
  \normalbaselineskip=11pt
  \setbox\strutbox=\hbox{\vrule height8pt depth3pt width0pt}
  \let\sc=\sevenrm \let\big=\ninebig \normalbaselines\rm}
\gdef\eightpoint{\def\rm{\fam0\eightrm}
  \textfont0=\eightrm \scriptfont0=\sixrm \scriptscriptfont0=\fiverm
  \textfont1=\eighti \scriptfont1=\sixi \scriptscriptfont1=\fivei
  \textfont2=\eightsy \scriptfont2=\sixsy \scriptscriptfont2=\fivesy
  \textfont3=\tenex \scriptfont3=\tenex \scriptscriptfont3=\tenex
  \def\mcal{\fam2 \eightsy}  \def\mmit{\fam1 \eighti}
  \textfont\itfam=\eightit \def\it{\fam\itfam\eightit}
  \textfont\slfam=\eightsl \def\sl{\fam\slfam\eightsl}
  \textfont\ttfam=\eighttt \scriptfont\ttfam=\eighttt
  \scriptscriptfont\ttfam=\eighttt \def\tt{\fam\ttfam\eighttt}
  \textfont\bffam=\eightbf \scriptfont\bffam=\sixbf
  \scriptscriptfont\bffam=\fivebf \def\bf{\fam\bffam\eightbf}
     \ifx\arisposta\amsrisposta   \ifnum\contaeuler=1
  \textfont\eufmfam=\eighteufm \scriptfont\eufmfam=\sixeufm
  \scriptscriptfont\eufmfam=\fiveeufm \def\eufm{\fam\eufmfam\eighteufm}
  \textfont\eufbfam=\eighteufb \scriptfont\eufbfam=\sixeufb
  \scriptscriptfont\eufbfam=\fiveeufb \def\eufb{\fam\eufbfam\eighteufb}
  \def\eurm{\eighteurm} \def\eurb{\eighteurb} \def\eusm{\eighteusm}
  \def\eusb{\eighteusb}       \fi    \ifnum\contaams=1
  \textfont\msamfam=\eightmsam \scriptfont\msamfam=\sixmsam
  \scriptscriptfont\msamfam=\fivemsam \def\msam{\fam\msamfam\eightmsam}
  \textfont\msbmfam=\eightmsbm \scriptfont\msbmfam=\sixmsbm
  \scriptscriptfont\msbmfam=\fivemsbm \def\msbm{\fam\msbmfam\eightmsbm}
     \fi       \ifnum\contacyrill=1     \def\cyrill{\eightwncyr}
  \def\cyrilb{\eightwncyb}  \def\cyrili{\eightwncyi}         \fi
  \textfont3=\eightex \scriptfont3=\sevenex \scriptscriptfont3=\sevenex
  \def\cmmib{\fam\cmmibfam\eightcmmib}  \textfont\cmmibfam=\eightcmmib
  \scriptfont\cmmibfam=\sixcmmib \scriptscriptfont\cmmibfam=\fivecmmib
  \def\cmbsy{\fam\cmbsyfam\eightcmbsy}  \textfont\cmbsyfam=\eightcmbsy
  \scriptfont\cmbsyfam=\sixcmbsy \scriptscriptfont\cmbsyfam=\fivecmbsy
  \def\cmcsc{\fam\cmcscfam\eightcmcsc} \scriptfont\cmcscfam=\eightcmcsc
  \textfont\cmcscfam=\eightcmcsc \scriptscriptfont\cmcscfam=\eightcmcsc
     \fi             \tt \ttglue=.5em plus.25em minus.15em
  \normalbaselineskip=9pt
  \setbox\strutbox=\hbox{\vrule height7pt depth2pt width0pt}
  \let\sc=\sixrm \let\big=\eightbig \normalbaselines\rm }
\gdef\tenbig#1{{\hbox{$\left#1\vbox to8.5pt{}\right.\n@space$}}}
\gdef\ninebig#1{{\hbox{$\textfont0=\tenrm\textfont2=\tensy
   \left#1\vbox to7.25pt{}\right.\n@space$}}}
\gdef\eightbig#1{{\hbox{$\textfont0=\ninerm\textfont2=\ninesy
   \left#1\vbox to6.5pt{}\right.\n@space$}}}
\def\alternativefont#1#2{\ifx\arisposta\amsrisposta \relax \else
\xdef#1{#2} \fi}
\global\contaeuler=0 \global\contacyrill=0 \global\contaams=0
%
%
%
%
\newbox\fotlinebb \newbox\hedlinebb \newbox\leftcolumn
\gdef\makeheadline{\vbox to 0pt{\vskip-22.5pt
     \fullline{\vbox to8.5pt{}\the\headline}\vss}\nointerlineskip}
\gdef\makehedlinebb{\vbox to 0pt{\vskip-22.5pt
     \fullline{\vbox to8.5pt{}\copy\hedlinebb\hfil
     \line{\hfill\the\headline\hfill}}\vss} \nointerlineskip}
\gdef\makefootline{\baselineskip=24pt \fullline{\the\footline}}
\gdef\makefotlinebb{\baselineskip=24pt
    \fullline{\copy\fotlinebb\hfil\line{\hfill\the\footline\hfill}}}
\gdef\doubleformat{\shipout\vbox{\Landspec\makehedlinebb
     \fullline{\box\leftcolumn\hfil\columnbox}\makefotlinebb}
     \advancepageno}
\gdef\columnbox{\leftline{\pagebody}}
\gdef\line#1{\hbox to\hsize{\hskip\leftskip#1\hskip\rightskip}}
\gdef\fullline#1{\hbox to\fullhsize{\hskip\leftskip{#1}%
\hskip\rightskip}}
\gdef\footnote#1{\let\@sf=\empty
         \ifhmode\edef\#sf{\spacefactor=\the\spacefactor}\/\fi
         #1\@sf\vfootnote{#1}}
\gdef\vfootnote#1{\insert\footins\bgroup
         \ifnum\dimnota=1  \eightpoint\fi
         \ifnum\dimnota=2  \ninepoint\fi
         \ifnum\dimnota=0  \tenpoint\fi
         \interlinepenalty=\interfootnotelinepenalty
         \splittopskip=\ht\strutbox
         \splitmaxdepth=\dp\strutbox \floatingpenalty=20000
         \leftskip=\oldssposta \rightskip=\olddsposta
         \spaceskip=0pt \xspaceskip=0pt
         \ifnum\sinnota=0   \textindent{#1}\fi
         \ifnum\sinnota=1   \item{#1}\fi
         \footstrut\futurelet\next\fo@t}
\gdef\fo@t{\ifcat\bgroup\noexpand\next \let\next\f@@t
             \else\let\next\f@t\fi \next}
\gdef\f@@t{\bgroup\aftergroup\@foot\let\next}
\gdef\f@t#1{#1\@foot} \gdef\@foot{\strut\egroup}
\gdef\footstrut{\vbox to\splittopskip{}}
\skip\footins=\bigskipamount
\count\footins=1000  \dimen\footins=8in
\catcode`@=12
\tenpoint
\ifnum\unoduecol=1 \hsize=\tothsize   \fullhsize=\tothsize \fi
\ifnum\unoduecol=2 \hsize=\collhsize  \fullhsize=\tothsize \fi
\global\let\lrcol=L      \ifnum\unoduecol=1
\output{\plainoutput{\ifnum\tipbnota=2 \clearnmbnota\fi}} \fi
\ifnum\unoduecol=2 \output{\if L\lrcol
     \global\setbox\leftcolumn=\columnbox
     \global\setbox\fotlinebb=\line{\hfill\the\footline\hfill}
     \global\setbox\hedlinebb=\line{\hfill\the\headline\hfill}
     \advancepageno  \global\let\lrcol=R
     \else  \doubleformat \global\let\lrcol=L \fi
     \ifnum\outputpenalty>-20000 \else\dosupereject\fi
     \ifnum\tipbnota=2\clearnmbnota\fi }\fi
\def\ifdoublepage{\ifnum\unoduecol=2 }
\gdef\yespagenumbers{\footline={\hss\tenrm\folio\hss}}
\gdef\ciao{ \ifnum\fdefcontre=1 \endfdef\fi
     \par\vfill\supereject \ifnum\unoduecol=2
     \if R\lrcol  \headline={}\nopagenumbers\null\vfill\eject
     \fi\fi \end}

\newskip\olddsposta \newskip\oldssposta
\global\oldssposta=\leftskip \global\olddsposta=\rightskip

\def\filldots{\leaders\hbox to 1em{\hss.\hss}\hfill}
\def\inquadrb#1 {\vbox {\hrule  \hbox{\vrule \vbox {\vskip .2cm
    \hbox {\ #1\ } \vskip .2cm } \vrule  }  \hrule} }
 \def\newline{\hfil\break}
\def\jump{\vskip\baselineskip} \newskip\iinnffrr
\def\sjump{\iinnffrr=\baselineskip
          \divide\iinnffrr by 2 \vskip\iinnffrr}
\def\bjump{\vskip\baselineskip \vskip\baselineskip}
\newcount\nmbnota  \def\clearnmbnota{\global\nmbnota=0}
\newcount\tipbnota \def\letterfootnote{\global\tipbnota=1}

\def\note#1{\global\advance\nmbnota by 1 \ifnum\tipbnota=1
    \footnote{$^{\rm\nttlett}$}{#1} \else {\ifnum\tipbnota=2
    \footnote{$^{\nttsymb}$}{#1}
    \else\footnote{$^{\the\nmbnota}$}{#1}\fi}\fi}
\def\nttlett{\ifcase\nmbnota \or a\or b\or c\or d\or e\or f\or
g\or h\or i\or j\or k\or l\or m\or n\or o\or p\or q\or r\or
s\or t\or u\or v\or w\or y\or x\or z\fi}
\def\nttsymb{\ifcase\nmbnota \or\dag\or\sharp\or\ddag\or\star\or
\natural\or\flat\or\clubsuit\or\diamondsuit\or\heartsuit
\or\spadesuit\fi}   \clearnmbnota
\def\numberfootnote{\global\tipbnota=0} \numberfootnote
\def\setnote#1{\expandafter\xdef\csname#1\endcsname{
\ifnum\tipbnota=1 {\rm\nttlett} \else {\ifnum\tipbnota=2
{\nttsymb} \else \the\nmbnota\fi}\fi} }
\newcount\nbmfig  \def\clearnbmfig{\global\nbmfig=0}
\gdef\figure{\global\advance\nbmfig by 1
      {\rm fig. \the\nbmfig}}   \clearnbmfig
\def\setfig#1{\expandafter\xdef\csname#1\endcsname{fig. \the\nbmfig}}
 \def\endformula{\eqno\numero $$}
 \def\efr{\endformula}
\newcount\frmcount \def\clearfrmcount{\global\frmcount=0}
\def\numero{\global\advance\frmcount by 1   \ifnum\indappcount=0
  {\ifnum\cpcount <1 {\hbox{\rm (\the\frmcount )}}  \else
  {\hbox{\rm (\the\cpcount .\the\frmcount )}} \fi}  \else
  {\hbox{\rm (\applett .\the\frmcount )}} \fi}
\def\nameformula#1{\global\advance\frmcount by 1%
\ifnum\draftnum=0  {\ifnum\indappcount=0%
{\ifnum\cpcount<1\xdef\spzzttrra{(\the\frmcount )}%
\else\xdef\spzzttrra{(\the\cpcount .\the\frmcount )}\fi}%
\else\xdef\spzzttrra{(\applett .\the\frmcount )}\fi}%
\else\xdef\spzzttrra{(#1)}\fi%
\expandafter\xdef\csname#1\endcsname{\spzzttrra}
\eqno \hbox{\rm\spzzttrra} $$}
\def\nfr{\nameformula}    
\def\nameali#1{\global\advance\frmcount by 1%
\ifnum\draftnum=0  {\ifnum\indappcount=0%
{\ifnum\cpcount<1\xdef\spzzttrra{(\the\frmcount )}%
\else\xdef\spzzttrra{(\the\cpcount .\the\frmcount )}\fi}%
\else\xdef\spzzttrra{(\applett .\the\frmcount )}\fi}%
\else\xdef\spzzttrra{(#1)}\fi%
\expandafter\xdef\csname#1\endcsname{\spzzttrra}
  \hbox{\rm\spzzttrra} }      \clearfrmcount
\newcount\cpcount \def\clearcpcount{\global\cpcount=0}
\newcount\subcpcount \def\clearsubcpcount{\global\subcpcount=0}
\newcount\appcount \def\clearappcount{\global\appcount=0}
\newcount\indappcount \def\clearindappcount{\indappcount=0}
\newcount\sottoparcount 

\def\applett{\ifcase\appcount  \or {A}\or {B}\or {C}\or
{D}\or {E}\or {F}\or {G}\or {H}\or {I}\or {J}\or {K}\or {L}\or
{M}\or {N}\or {O}\or {P}\or {Q}\or {R}\or {S}\or {T}\or {U}\or
{V}\or {W}\or {X}\or {Y}\or {Z}\fi    \ifnum\appcount<0
\immediate\write16 {Panda ERROR - Appendix: counter "appcount"
out of range}\fi  \ifnum\appcount>26  \immediate\write16 {Panda
ERROR - Appendix: counter "appcount" out of range}\fi}
\clearappcount  \clearindappcount \newcount\connttrre
\def\clearconnttrre{\global\connttrre=0} \newcount\countref
\def\clearcountref{\global\countref=0} \clearcountref
\def\chapter#1{\global\advance\cpcount by 1 \clearfrmcount
                 \goodbreak\null\vbox{\jump\nobreak
                 \clearsubcpcount\clearindappcount
                 \itemitem{\ttaarr\the\cpcount .\qquad}{\ttaarr #1}
                 \par\nobreak\jump\sjump}\nobreak}
\def\section#1{\global\advance\subcpcount by 1 \goodbreak\null
               \vbox{\sjump\nobreak\ifnum\indappcount=0
                 {\ifnum\cpcount=0 {\itemitem{\ppaarr
               .\the\subcpcount\quad\enskip\ }{\ppaarr #1}\par} \else
                 {\itemitem{\ppaarr\the\cpcount .\the\subcpcount\quad
                  \enskip\ }{\ppaarr #1} \par}  \fi}
                \else{\itemitem{\ppaarr\applett .\the\subcpcount\quad
                 \enskip\ }{\ppaarr #1}\par}\fi\nobreak\jump}\nobreak}
\clearsubcpcount
\def\appendix#1{\global\advance\appcount by 1 \clearfrmcount
                  \goodbreak\null\vbox{\jump\nobreak
                  \global\advance\indappcount by 1 \clearsubcpcount
          \itemitem{ }{\hskip-40pt\ttaarr Appendix\ #1}
             \nobreak\jump\sjump}\nobreak}
\clearappcount \clearindappcount
\def\references{\goodbreak\null\vbox{\jump\nobreak
   \itemitem{}{\ttaarr References} \nobreak\jump\sjump}\nobreak}

\clearcpcount\clearcountref

\def\setchap#1{\ifnum\indappcount=0{\ifnum\subcpcount=0%
\xdef\spzzttrra{\the\cpcount}%
\else\xdef\spzzttrra{\the\cpcount .\the\subcpcount}\fi}
\else{\ifnum\subcpcount=0 \xdef\spzzttrra{\applett}%
\else\xdef\spzzttrra{\applett .\the\subcpcount}\fi}\fi
\expandafter\xdef\csname#1\endcsname{\spzzttrra}}
\newcount\draftnum \newcount\ppora   \newcount\ppminuti
\global\ppora=\time   \global\ppminuti=\time
\global\divide\ppora by 60  \draftnum=\ppora
\multiply\draftnum by 60    \global\advance\ppminuti by -\draftnum
\def\droggi{\number\day /\number\month /\number\year\ \the\ppora
:\the\ppminuti}     \global\draftnum=0
\def\draftcomment#1{\ifnum\draftnum=0 \relax \else
{\ {\bf ***}\ #1\ {\bf ***}\ }\fi} 
%
%
\catcode`@=11
\gdef\Ref#1{\expandafter\ifx\csname @rrxx@#1\endcsname\relax%
{\global\advance\countref by 1    \ifnum\countref>200
\immediate\write16 {Panda ERROR - Ref: maximum number of references
exceeded}  \expandafter\xdef\csname @rrxx@#1\endcsname{0}\else
\expandafter\xdef\csname @rrxx@#1\endcsname{\the\countref}\fi}\fi
\ifnum\draftnum=0 \csname @rrxx@#1\endcsname \else#1\fi}
\gdef\beginref{\ifnum\draftnum=0  \gdef\Rref{\fairef}
\gdef\endref{\scriviref} \else\relax\fi
\ifx\risposta\mplarisposta \ninepoint \fi
\parskip 2pt plus.2pt \baselineskip=12pt}
\def\Reflab#1{[#1]} \gdef\Rref#1#2{\item{\Reflab{#1}}{#2}}
\gdef\endref{\relax}  \newcount\conttemp
\gdef\fairef#1#2{\expandafter\ifx\csname @rrxx@#1\endcsname\relax
{\global\conttemp=0 \immediate\write16 {Panda ERROR - Ref: reference
[#1] undefined}} \else
{\global\conttemp=\csname @rrxx@#1\endcsname } \fi
\global\advance\conttemp by 50  \global\setbox\conttemp=\hbox{#2} }
\gdef\scriviref{\clearconnttrre\conttemp=50
\loop\ifnum\connttrre<\countref \advance\conttemp by 1
\advance\connttrre by 1
\item{\Reflab{\the\connttrre}}{\unhcopy\conttemp} \repeat}
\clearcountref \clearconnttrre
\catcode`@=12
\ifx\risposta\mplarisposta \def\Reflab#1{#1.} \letterfootnote \fi

\def\slashchar#1{\setbox0=\hbox{$#1$} \dimen0=\wd0
     \setbox1=\hbox{/} \dimen1=\wd1 \ifdim\dimen0>\dimen1
      \rlap{\hbox to \dimen0{\hfil/\hfil}} #1 \else
      \rlap{\hbox to \dimen1{\hfil$#1$\hfil}} / \fi}
\ifx\oldchi\undefined \let\oldchi=\chi
  \def\cchi{{\raise 1pt\hbox{$\oldchi$}}} \let\chi=\cchi \fi

\def\frac#1#2{{\textstyle{#1 \over #2}}}

\def\half{\ifinner {\scriptstyle {1 \over 2}}\else {1 \over 2} \fi}
  \def\ket#1{\vert#1\rangle}

\def\simge{\rlap{\raise 2pt \hbox{$>$}}{\lower 2pt \hbox{$\sim$}}}
\def\simle{\rlap{\raise 2pt \hbox{$<$}}{\lower 2pt \hbox{$\sim$}}}

\def\vbig#1#2{{\vbigd@men=#2\divide\vbigd@men by 2%
\hbox{$\left#1\vbox to \vbigd@men{}\right.\n@space$}}}

%
%
\newcount\fdefcontre \newcount\fdefcount \newcount\indcount
\newread\filefdef  \newread\fileftmp  \newwrite\filefdef
\newwrite\fileftmp     \def\strip#1*.A {#1}
\def\futuredef#1{\beginfdef
\expandafter\ifx\csname#1\endcsname\relax%
{\immediate\write\fileftmp {#1*.A}
\immediate\write16 {Panda Warning - fdef: macro "#1" on page
\the\pageno \space undefined}
\ifnum\draftnum=0 \expandafter\xdef\csname#1\endcsname{(?)}
\else \expandafter\xdef\csname#1\endcsname{(#1)} \fi
\global\advance\fdefcount by 1}\fi   \csname#1\endcsname}

\def\beginfdef{\ifnum\fdefcontre=0
\immediate\openin\filefdef \jobname.fdef
\immediate\openout\fileftmp \jobname.ftmp
\global\fdefcontre=1  \ifeof\filefdef \immediate\write16 {Panda
WARNING - fdef: file \jobname.fdef not found, run TeX again}
\else \immediate\read\filefdef to\spzzttrra
\global\advance\fdefcount by \spzzttrra
\indcount=0      \loop\ifnum\indcount<\fdefcount
\advance\indcount by 1   \immediate\read\filefdef to\spezttrra
\immediate\read\filefdef to\sppzttrra
\edef\spzzttrra{\expandafter\strip\spezttrra}
\immediate\write\fileftmp {\spzzttrra *.A}
\expandafter\xdef\csname\spzzttrra\endcsname{\sppzttrra}
\repeat \fi \immediate\closein\filefdef \fi}
\def\endfdef{\immediate\closeout\fileftmp   \ifnum\fdefcount>0
\immediate\openin\fileftmp \jobname.ftmp
\immediate\openout\filefdef \jobname.fdef
\immediate\write\filefdef {\the\fdefcount}   \indcount=0
\loop\ifnum\indcount<\fdefcount    \advance\indcount by 1
\immediate\read\fileftmp to\spezttrra
\edef\spzzttrra{\expandafter\strip\spezttrra}
\immediate\write\filefdef{\spzzttrra *.A}
\edef\spezttrra{\string{\csname\spzzttrra\endcsname\string}}
\iwritel\filefdef{\spezttrra}
\repeat  \immediate\closein\fileftmp \immediate\closeout\filefdef
\immediate\write16 {Panda Warning - fdef: Label(s) may have changed,
re-run TeX to get them right}\fi}
\def\iwritel#1#2{\newlinechar=-1
{\newlinechar=`\ \immediate\write#1{#2}}\newlinechar=-1}
\global\fdefcontre=0 \global\fdefcount=0 \global\indcount=0
%
%
\null
%
%
%
%

\input psfig
%
\loadamsmath
\loadeuler
\mathchardef\bbalpha="710B
\mathchardef\bbbeta="710C
\mathchardef\bbgamma="710D
\mathchardef\bbomega="7121

\def\xn{{\theta\over 2\pi i}}
\def\ta{\theta}

\def\ep{\epsilon}
\def\th{\theta}
\def\kg{\Gamma}

\pageno=0\baselineskip=14pt
\nopagenumbers{
\line{\hfill SWAT/116}
\line{\hfill\tt hep-th/9606116}
\line{\hfill June 1996}
\ifdoublepage \bjump\bjump\bjump\bjump\else\vfill\fi
\centerline{\capsone The $N=1$ supersymmetric bootstrap and Lie algebras}
\bjump\bjump
\centerline{\scaps Timothy J. Hollowood and Evangelos Mavrikis} 
\sjump
\sjump
\centerline{\sl Department of Physics, University of Wales Swansea,}
\centerline{\sl Singleton Park, Swansea SA2 8PP, U.K.}
\centerline{\tt t.hollowood, e.mavrikis @swansea.ac.uk} 
\bjump\bjump\bjump
\ifdoublepage
\vfill
\noindent
\line{SWAT/\hfill}
\line{199\hfill}
\eject\null\vfill\fi
\centerline{\capsone ABSTRACT}\sjump
The bootstrap programme for finding exact S-matrices of integrable
quantum field theories with N=1 supersymmetry 
is investigated. New solutions are found which have the same fusing data
as bosonic theories related to the classical 
affine Lie algebras. When the states
correspond to a spinor spot of the Dynkin diagram they are kinks which
carry a non-zero topological charge. Using these results, the S-matrices
of the supersymmetric O($2n$) sigma model and sine-Gordon model can be
described in a uniform way.
\sjump\vfill
\ifdoublepage \else
\eject}
\yespagenumbers\pageno=1
%
%

\chapter{Introduction}

If a field theory in two-dimensions is integrable then its S-matrix
factorizes and the bootstrap programme becomes manageable. Many field
theories have been solved exactly by applying the axioms of S-matrix theory
in tandem with the bootstrap. It is fascinating that the known solutions of 
the bootstrap equations are related to (affine) Lie algebras. The
bootstrap is characterized by a set of data known as the fusing angles
which seem to have a rather universal character since they
re-appear in many different theories. 

The question that we set ourselves in this paper is whether there is a similar
picture for theories with supersymmetry. In particular, we consider the case
of $N=1$ supersymmetry. In the presence of supersymmetry, the
bootstrap equations become more complicated since in 
internal lines one must sum over all states in a
super-multiplet. Nevertheless, 
the solutions which we find have a very characteristic form. Let us suppose
that the $i^{\rm th}$ multiplet is labelled $\ket{\xi_i,A_i(\th)}$, where
$A_i$ specifies the supersymmetric quantum numbers, i.e. usually boson
or fermion, but sometimes more complicated `kink' representations appear,
and $\xi_i$ specifies any additional quantum numbers needed to label the
states.\note{$\th$ is the rapidity of the state.} 
Our S-matrix elements have the form of an ansatz made by Schoutens
[\Ref{S1}] in which there is no mixing
between the supersymmetric and internal quantum numbers. This mean
that the S-matrices have the form of a product
$$
tilde S_{\xi_i\xi_j\rightarrow\xi'_j\xi'_i}(\th)
S_{A_iA_j\rightarrow A'_jA'_i}(\th)=_{\ \rm out}
\langle\xi'_j,A'_j(\th_2)\xi'_i,A'_i(\th_1)\vert
\xi_i,A_i(\th_1)\xi_j,A_j(\th_2)\rangle_{\rm in},
\nfr{FAC}
where $\th=\th_1-\th_2$. In this ansatz, each of the factors separately
satisfies the axioms of S-matrix theory: unitarity and
crossing. Moreover, 
$tilde S(\th)$ is an S-matrix of of purely bosonic theory in its
own right and hence satisfies the bootstrap equations with some
characteristic set of fusing angles $u_{ij}^k$.\note{In this notation
the fusion occurs at the imarginary rapidity difference $iu_{ij}^k$.}
This implies that the
supersymmetric part $S(\th)$ has to satisfy the same bootstrap equations,
i.e. with the same fusing angles; however, since the bound-state pole is
already present in the factor $tilde S(\th)$ the supersymmetric factor satisfies
the bootstrap equations passively, in the sense that it does not have a pole
at the fusing rapidity $\th=iu_{ij}^k$. In fact, the supersymmetric S-matrix
factors that we will construct introduce no additional poles onto the
physical strip and hence do not introduce any additional
bound-states. This means that the spectrum of states and their fusings
follow exactly those of the bosonic theory.

The formalism of supersymmetric factorizable S-matrices was considered
some time ago; in particular we refer to the work of Bernard and
LeClair [\Ref{BL}], Ahn [\Ref{A1}] and Schoutens
[\Ref{S1}]. What was missing from this early work
were complete solutions to the bootstrap programme. We shall find that
in certain cases, to find consistent solutions of the bootstrap
equations we shall have to consider supersymmetric states which carry
topological charge. It transpires that these representations are not
carried by particle states, rather they are carried by
kinks. 

\chapter{Minimal supersymmetric S-matrix blocks}

In this section we review the construction of the basic S-matrix
building blocks describing the interactions of
supersymmetric particles and kinks. 
The question of how to put these blocks together to form a consistent
S-matrix for which the bootstrap closes, will be treated in
later sections. 

Supersymmetry is realized on a super-multiplet of states $A_i$, where
$i$ labels the multiplet
$$
{\cal  Q}\ket{A_i(\ta)}=\sqrt{m_i}e^{\ta / 2}{Q}
\ket{A_i(\ta)},
\qquad{\bar{\cal  Q}}\ket{A_i(\ta)}=\sqrt{m_i}e^{-\ta / 2}\bar
Q\ket{A_i(\ta)},
\nfr{sug}
where $\th$ is the rapidity of the state and $Q$ and $\bar Q$ are
matrices which act on the states of the super-multiplet and which satisfy
$$
Q^2=1,\qquad\bar{Q}^2=1.
\efr
The action of supersymmetry on multiple states involves braiding factors:
$$
\eqalign{
& {\cal Q} \ket{A_1(\ta _1) A_2(\ta _2)\cdots A_N(\ta _N)} \cr
& =\sum_{k=1}^{N}\sqrt{m_k} ~ e^{\ta _k /2}\ket{(Q_L A_1(\ta _1)) 
\cdots(Q_L  A_{k-1}(\ta _{k-1})) (QA_k(\ta _k))A_{k+1}
(\ta _{k+1})\cdots A_N(\ta _N)}.\cr} 
\nfr{tr1}
Similarly for $\bar{\cal Q}$
$$
\eqalign{
& \bar{{\cal Q} }\ket{A_1(\ta _1) A_2(\ta _2)\cdots A_N(\ta _N)} \cr
&=\sum_{k=1}^{N}\sqrt{m_k} ~ e^{-\ta _k /2}\ket{(Q_L A_1(\ta
_1))\cdots(Q_L  A_{k-1}(\ta _{k-1})) (\bar{Q}A_k(\ta _k))A_{k+1}(\ta
_{k+1})\cdots A_N(\ta _N)}. \cr} 
\efr
In the above, $Q_L$ is the fermion parity operator. 
In particular, for two states we have
$$
\eqalign{
{\cal Q}\ket{A_1(\ta _1) A_2(\ta _2)}&=\sqrt{m_1}~e^{\th _1 /2} 
Q_1\ket{A_1(\ta _1) A_2(\ta _2)}+\sqrt{m_2}~e^{\th _2 /2}
Q_2\ket{A_1(\ta _1)A_2(\ta _2)},\cr
\bar{{\cal Q}}\ket{A_1(\ta _1) A_2(\ta _2)}&=\sqrt{m_1}~e^{-\th _1 /2}
\bar{Q}_1\ket{A_1(\ta _1) A_2(\ta _2)}+\sqrt{m_2}~e^{-\th _2 /2}
\bar{Q}_2\ket{A_1(\ta _1) A_2(\ta _2)},\cr}
\efr
where $Q_1=Q\otimes I$, $Q_2=Q_L\otimes Q$, $\bar Q_1=Q\otimes I$ and
$\bar Q_2=Q_L\otimes\bar Q$.

A factorizable S-matrix is specified completely by the two-body
S-matrix elements which are defined as follows:
$$
\ket{A_i(\th_1)A_j(\th_2)}_{\rm in}=
\sum_{A'_iA'_j}S_{A_iA_j\rightarrow A'_jA'_i}
(\th_1-\th_2)\ket{A'_j(\th_2)A'_i(\th_1)}_{\rm out}.
\efr

\section{The scattering of particles}

The supermultiplets are doublets
containing a boson $\ket{\phi} $ and a fermion $\ket{\psi}$.
A suitable representation for the supercharges, in the basis $\{
\phi,\psi\}$, is [\Ref{S1}]
$$
Q=\pmatrix {0&\ep \cr \ep^{*} &0 \cr },~~~ \bar{Q}=\pmatrix {0&\ep ^* 
\cr \ep  &0 \cr }, 
\nfr{ors}
where $\ep=\exp(i\pi/4)$. The fermionic parity operator $Q_L$ is 
$$
Q_L=\pmatrix{1&0 \cr 0&-1\cr}.
\efr
It can be easily checked that this representation of supersymmetry
carries zero topological charge: $T=\half\{\bar{Q},Q\}=0$.

Here we describe the S-matrix building blocks constructed
by Schoutens [\Ref{S1}] which describe the scattering of two
such particles supermultiplets of masses $m_1$ and $m_2$. 
For a supersymmetric theory the action of
supersymmetry commutes with the S-matrix:\note{The subscript P
indicates that these S-matrices describe the
scattering of particle states, as opposed to kinks described in the
next subsection.}
$$
{\cal Q}S_{\rm P}(\th)=S_{\rm P}(\th){\cal Q},\qquad\bar{{\cal
Q}}S_{\rm P}(\th)=S_{\rm P}(\th)\bar{{\cal Q}}. 
\nfr{sucom}
We also require that the Yang-Baxter equation is satisfied which
ensures that the S-matrix is 
factorizable.\note{The Yang-Baxter equation is non-trivial
in this context because there are non-diagonal processes; for instance
the boson can reflect off the fermion.}
The general solution of Schoutens is then fixed up to an
overall scalar function $G(\th)$ and a constant $\alpha$.
$$\eqalign{
&S_{\rm P}(\th)=G(\th){1\over2i}(Q_1-Q_2)(\bar Q_1-\bar Q_2)\cr
&+\alpha F(\th)\left(1-{\rm
tanh}\left({\ta+\log(m_1/m_2)\over4}\right)Q_1Q_2\right)\left(
1+{\rm
tanh}\left({\ta-\log(m_1/m_2)\over4}\right)\bar Q_1\bar Q_2\right),\cr}
\nfr{gma}
where
$$
F(\ta)={m_1+m_2+2\sqrt{m_1m_2}\cosh(\ta/2)\over
2i\sinh\ta}G(\ta).
\efr
The constant $\alpha$  measures the strength of Bose-Fermi mixing 
interactions.

The function $G(\th)$ is determined by imposing unitarity
and crossing symmetry. A minimal solution for the case when the
particles are self-conjugate was found in [\Ref{S1}]. We shall
write down explicit expressions for these factors in section 3.1.

\section{The scattering of kinks}

If there are topologically charged states in the theory, then the supersymmetry algebra
is modified by central charges [\Ref{WO}]. In general, states
which carry topological charges are kinks $K_{ab}(\th)$ which
interpolate between two vacua $a,b\in\Gamma$, where $\Gamma$ is the
set of vacua. It is important to realize
that kink states are {\it not\/} in general equivalent to a set of
particle states, due to fact that multi-kink states must respect an
adjacency condition, i.e. $|\cdots
K_{ab}(\th_1)K_{cd}(\th_2)\cdots\rangle$ requires $b=c$. 

An S-matrix with supersymmetry can be associated 
to a system whose fundamental excitations are four 
kinks, interpolating between three vacua, labelled by $0$, $\half$, $1$.
In the basis $\{
K_{0\half},K_{1\half},K_{\half 0},K_{\half1} \}$, and appropriate
representation of supersymmetry is
$$
Q=\pmatrix{0&i&0&0 \cr -i&0&0&0 \cr 0&0&1&0 \cr 0&0&0&-1},~~
\bar{Q}=\pmatrix{0&i&0&0 \cr -i&0&0&0 \cr 0&0&-1&0 \cr
0&0&0&1},~~Q_L=\pmatrix{0&1&0&0 \cr 1&0&0&0 \cr 
0&0&0&1\cr 0&0&1&0}.
\nfr{Sak}
The topological charges are
$$
T={1\over2}\{Q,\bar{Q}\}=\pmatrix{1&0&0&0
\cr 0&1&0&0 \cr 0&0&-1&0\cr 0&0&0&-1\cr},
\efr
so $K_{0\half}$ and $K_{1\half}$ have $T=1$, and 
$K_{\half 0}$ and $K_{\half 1}$ have $T=-1$.

The S-matrix describing the scattering of the kinks can be written as
[\Ref{S1}] 
$$
S_{\rm K}(\th)=K(\th)(\cosh(\gamma \th)-\sinh(\gamma \th)Q_1\bar{Q}_1)
(\cosh(\th /4)-\sinh(\th /4)Q_1 Q_2),
\efr
where $\gamma=\log2/2\pi i$. We will denote the explicit S-matrix
elements for $K_{ab}(\th_1)+K_{bc}(\th_2)\rightarrow
K_{ad}(\th_2)+K_{dc}(\th_1)$ as
$$
S\left(\left.\matrix{a&d\cr b&c\cr}\right\vert\th_1-\th_2\right).
\efr
Explicitly, the non-zero elements are
$$\eqalign{
S\left(\left.\matrix{0&\half\cr\half &0\cr}
\right\vert\th\right)&= S\left(\left.
\matrix{1&\half\cr\half &1\cr}\right\vert\th\right)=K(\th)2^{(i\pi-\th)/2\pi
i}\cos\left({\th\over4i}-{\pi\over4}\right)\cr
S\left(\left.\matrix{\half&0\cr0 &\half\cr}\right\vert\th\right)&= 
S\left(\left.
\matrix{\half&1\cr 1&\half\cr}\right\vert\th\right)=K(\th)2^{\th/2\pi
i}\cos\left({\th\over4i}\right)\cr
S\left(\left.\matrix{0&\half\cr\half &1\cr}\right\vert\th\right)&= 
S\left(\left.
\matrix{1&\half\cr \half&0\cr}\right\vert\th\right)=K(\th)2^{(i\pi-\th)/2\pi
i}\cos\left({\th\over4i}+{\pi\over4}\right)\cr
S\left(\left.\matrix{\half&1\cr0 &\half\cr}\right\vert\th\right)&= 
S\left(\left.
\matrix{\half&0\cr 1&\half\cr}\right\vert\th\right)=K(\th)2^{\th/2\pi
i}\cos\left({\th\over4i}-{\pi\over2}\right)\cr}
\efr
The scalar scalar function $K(\th)$ is determined by crossing symmetry and
unitarity. The minimal solution for $K(\th)$ is
$$
K(\th)={1\over\sqrt\pi}
\prod_{k=1}^{\infty}{ \kg(k-\half +\th /2\pi i)
\kg (k-\th /2\pi i)\over\kg (k+\th /2\pi i)\kg(k+\half -\th /2\pi i)}.
\nfr{chm}

This S-matrix was first written down by Zamolodchikov [\Ref{zam5}] to describe
the perturbation of the $c=7/10$ superconformal minimal model by an operator of
dimension $6/5$. Subsequently it was realized that the
S-matrix elements are related to a solution of the Yang-Baxter equation
associated to the quantum group $U_q(su(2))$ [\Ref{ABL}]. 

\chapter{The fusing procedure}

In the last section we described the construction of supersymmetric 
S-matrix blocks describing particles and kinks. In this section we
show how such blocks can be put together to describe supersymmetric
theories with rich spectra of states. 

As it stands the supersymmetric S-matrices that we have constructed in
the last section are `still-born' in the sense that they have no
bound-state poles on the physical strip. In order to rectify this
we can take, following Schoutens [\Ref{S1}], an ansatz which consists of a 
product of the S-matrix of purely bosonic theory
${S}'(\th)$ with the supersymmetric S-matrix blocks of the last
section which we denote as $S(\th)$. The
resulting S-matrix is then schematically of the form
$$
S_{\rm SUSY}(\th)={S}'(\th) \otimes {S}(\th).
\efr
This notation is short-hand for the product form displayed in \FAC.
These S-matrices 
describe a set of states carrying the quantum numbers of the bosonic
theory as well as forming either particle or kink supermultiplets.
An ansatz of this form manifestly satisfies unitarity and crossing symmetry,
because each of the factors does so separately. The only 
non-trivial consistency
conditions arise from the existence of bound-states due to poles in
the bosonic factor. By itself the bosonic part, of course, satisfies
the required bootstrap equations; however, what has to be checked is
that the supersymmetric part does not introduce any additional poles onto
the physical strip and satisfies the {\it same\/} bootstrap
equations, although only passively, in the sense that on its own it
does not have the required bound-state poles. 
This implies that in our
discussion in this section the expression for the supersymmetric part
of the bound-state wavefunction is given by
the limit of the two particle state, at the position of the pole in
the bosonic part, rather than the residue. Let us suppose that the
fusing angles of the bosonic theory are $u_{ij}^k$.
The bootstrap equations for $S(\th)$ can be summed up by the relation
$$
\vert A_i(\th+i\bar u_{i\bar k}^{\bar j})
A_j(\th-i\bar u_{j\bar k}^{\bar i})\rangle=\sum_{A_k}f_{A_i
A_j}^{A_k}\vert A_k(\th)\rangle,
\efr
where the sum is over all states in the $k^{\rm th}$ super-multiplet. In
the above we have defined $\bar u=\pi-u$ and $\bar j$ is the charge
conjugate multiplet of $j$.
The coupling constants are related to the S-matrix evaluated at the
fusing rapidity:
$$
S_{A_iA_j\rightarrow A'_jA'_i}(iu_{ij}^k)=\sum_{A_k}
f_{A_iA_j}^{A_k}\left(f_{A'_iA'_j}^{A_k}\right)^*.
\nfr{CCSM}
Notice that in the above formula we do not relate the coupling
constants to the residue of the S-matrix as one does for the bosonic
part of the S-matrix, since $S(\th)$ has no pole.

Our task in this section is to show in a general context 
how the supersymmetric blocks solve the bootstrap equations with the same
fusing angles encountered in the bosonic theories. In the section
4  and 5 we go on to construct explicitly supersymmetric S-matrices
by consider particular bosonic factors $tilde S(\th)$ and their fusing structure.

\section{Fusing between particles}

In this section we consider the bound-states that can appear in the
S-matrix $S_{\rm P}(\th)$ describing the scattering of supersymmetric
particles. We denote the mass of the $a^{rm th}$ super-multiplet $m_a$.
If the bosonic part of the S-matrix provides a bound-state
pole at $\th=iu_{ab}^c$ in the scattering of $a$ with $b$ then the
bootstrap equations for the supersymmetric part of the S-matrix are
equivalent to defining the wavefunction of the bound-state as: 
$$
\eqalign{
\ket{\phi_c (\th)}&={1\over(f_{\phi \phi})_{ab}^c}
\ket{\phi _a(\th+i\bar{u}_{a\bar c}^{\bar b})
\phi_b(\th-i\bar{u}_{b\bar c}^{\bar a})}= 
{1\over(f_{\psi \psi})_{ab}^c}
\ket{\psi _a(\th+i\bar{u}_{a\bar c}^{\bar b})
\psi_b(\th-i\bar{u}_{b\bar c}^{\bar a})}  \cr 
\ket{\psi_c (\th)}&={1\over(f_{\phi \psi})_{ab}^c}
\ket{\phi _a(\th+i\bar{u}_{a\bar c}^{\bar b})
\psi_b(\th-i\bar{u}_{b\bar c}^{\bar a})} 
={1\over(f_{\psi \phi})_{ab}^c}
\ket{\psi _a(\th+i\bar{u}_{a\bar c}^{\bar b})
\phi_b(\th-i\bar{u}_{b\bar c}^{\bar a})}, \cr}
\nfr{dec}
where $\bar a$ denotes the charge conjugate of $a$ and the coupling
constants are related to the S-matrix elements via \CCSM\ from which
one deduces that
$$ 
\eqalign{
&(f_{\phi \phi}) _{ab}^c=\sqrt{S^{[ab]}_{\phi
\phi\rightarrow\phi\phi}(iu_{ab}^c}),\qquad(f_{\psi \psi})_{ab}^c=
\sqrt{S^{[ab]}_{\psi\psi\rightarrow\psi\psi}(iu_{ab}^c)} \cr
&(f_{\phi \psi})_{ab}^c=\sqrt{S^{[ab]}_{\phi\phi\rightarrow\psi
\psi}(iu_{ab}^c}),\qquad(f_{\psi \phi})_{ab}^c=
\sqrt{S^{[ab]}_{\psi\psi\rightarrow\phi\phi}(iu_{ab}^c)}.\cr}
\nfr{tired}
From this ift follows
$$
{(f_{\psi \psi})_{ab}^c\over(f_{\phi \phi})_{ab}^c}=\sqrt{m_a
+ m_b -m_c\over m_a + m_b +m_c},~ {(f_{\phi \psi})_{ab}^c\over
(f_{\phi\phi})_{ab}^c}=\sqrt{ m_b -m_a + m_c\over m_a + m_b
+m_c},~{(f_{\psi \phi})_{ab}^c\over(f_{\phi
\phi})_{ab}^c}=\sqrt{m_a -m_b + m_c\over m_a + m_b +m_c}. 
\efr
The definitions of these bound-states are consistent with the action of
supersymmetry [\Ref{S1}].

These equations are highly restrictive and determine the coupling
constant $\alpha$ [\Ref{S1}]:
$$
\alpha=-{\sin (u_{ab}^c)\over m_c}=-{\sqrt{4m_a^2m_b ^2-(m_a ^2
+m_b^2 -m_c^2)^2}\over2m_a m_b m_c}, 
\nfr{key}
which in turn
places a highly restrictive constraint on the
allowed fusing rules and mass spectrum in the theory. 
The spectrum is fixed up to a constant $H$:
$$
m_a=m\sin (a\pi /H),\ \ \ a=1,2,\ldots,n,
\nfr{gsp}
where the total number of particles $n$ will depend on $H$.
The allowed fusings depend upon whether the particles are
self-conjugate or not. In the former case we have 
$n=[H/2]$ and the only fusing angles which are consistent with the
mass spectrum (for generic values of $H$) are
$$
u_{ab}^{a+b}={(a+b)\pi\over H}\quad(a+b\leq n),\qquad
u_{ab}^{|b-a|}=\pi- {|b-a|\pi\over H}.
\nfr{fao}
For some theories considered in the next section, 
$H$ will be constrained to some integer value and there exist additional
fusing angles. Also in one class of solutions to the bootstrap, we shall
find the particles are not self-conjugate.
These points will be discussed fully in section 4.
For the present, we will work out the implicitions of
the fusing rules in \fao.

It is straightforward to show that the crucial 
condition \key\ is satisfied
for each of the fusings \fao\ and the value of the coupling
constant is fixed to be $\alpha^{-1}=-m$. The S-matrix elements
describing the scattering of particle supermultiplets 
are explicitly 
$$
\eqalign{
S_{\phi\phi
\rightarrow\phi\phi}^{[ab]}(\th)&=\left(1+{2\sin({a+b\over 2H}\pi)\cos(
{a-b\over 2H}\pi)\over \sin({\theta\over i})} \right)G^{[ab]}(\theta),\cr
S_{\psi \psi\rightarrow\psi\psi}^{[ab]}
(\theta)&=\left(-1+{2\sin({a+b\over 2H}\pi)\cos({a-b\over 2H}\pi)
\over\sin({\theta\over i})}\right)G^{[ab]}(\theta),\cr 
S_{\phi\phi\rightarrow\psi\psi}^{[ab]}(\theta)&=S_{\psi 
\psi\rightarrow\phi\phi}^{[ab]}(\theta)=
{\sqrt{\sin({a\pi\over H})\sin({b\pi\over H})}\over \cos({\theta\over 2i})}
G^{[ab]}(\theta),\cr
S_{\phi\psi\rightarrow\phi\psi}^{[ab]}(\theta)&=
S_{\psi\phi\rightarrow\psi\phi}^{[ab]}(\theta)=
{\sqrt{\sin({a\pi\over H})\sin({b\pi\over H})}\over 
\sin({\theta\over2i})}G^{[ab]}(\theta),\cr
S_{\phi\psi\rightarrow\psi\phi}^{[ab]}(\theta)&=
\left(1-{2\sin({a-b\over 2H}\pi)\cos(
{a+b\over 2H}\pi)\over \sin({\theta\over i})}
\right)G^{[ab]}(\theta), \cr
S_{\psi\phi\rightarrow\phi\psi}^{[ab]}(\theta)&=
\left(1+{2\sin({a-b\over 2H}\pi)\cos(
{a+b\over 2H}\pi)\over \sin({\theta\over i})}
\right)G^{[ab]}(\theta).\cr}
\nfr{rgo}

We must now verify that the S-matrix elements above satisfy the
bootstrap equations which follow from the fusing rules \fao.
First of all let us derive the expressions for the
unitarizing/crossing factors $G^{[ab]}(\th)$. We will assume for the
remainder of this section that the particles are self-conjugate.
The unitarity and crossing symmetry  constraints for the 
S-matrix elements \rgo\ require
$$
\eqalign{
G^{[ab]}(\th)G^{[ab]}(-\th)&={\sin^2({\th \over 2i})
\cos^2({\th \over 2i})\over \sin ({\th \over 2i}+{(a+b)\pi\over 
2H})\sin ({\th \over 2i}-{(a+b)\pi\over  2H})\cos ({\th
\over 2i}+{(a-b)\pi\over  2H})\cos ({\th \over 2i}-{(a-b)\pi\over 
2H})},\cr
G^{[ab]}(\th)&=G^{[ab]}(i\pi-\th).\cr}
\nfr{uac}
It is useful to rewrite the second equation above using the first as
$$ \eqalign {
G^{[ab]}&(i\pi +\th)G^{[ab]}(i\pi -\th) \cr
&= {\sin^2({\th \over 2i}) \cos^2({\th \over 2i})\over \sin
({\th \over 2i}+{(a+b)\pi\over  2H})\sin ({\th
\over 2i}-{(a+b)\pi\over  2H})\cos ({\th \over 2i}+{(a-b)\pi\over 
2H})\cos ({\th \over 2i}-{(a-b)\pi\over  2H})}.\cr}
\nfr{uncr}     
Solving the above system for $G^{[ab]}(\th)$ we find:
$$
\eqalign{ 
G^{[ab]}(\th)&=R^{[ab]}(\th)R^{[ab]}(i\pi-\th),\cr
R^{[ab]}(\th)&={1\over \kg(\xn)\kg(\xn+\half)}  \prod_{k=1}^{\infty}
{\kg(\xn +{a+b\over2H}+k-1)  \kg(\xn - {a+b\over2H} +k)\over \kg(\xn
+{a+b\over2H} +k-\half) \kg(\xn -{a+b\over2H}+k+ \half)} \cr  
&\qquad\times {\kg(\xn + {a-b\over2H} +k- \half ) \kg( \xn -{a-b\over2H}+k
-\half)\over  \kg( \xn +{a-b\over2H} +k) \kg( \xn -{a-b\over2H} +k)}.\cr}
\nfr{niania}
This solution follows from the general expression in [\Ref{S1}].
These expressions are not unique since they are subject to the usual CDD
ambiguities. However, they are the minimal solutions: the ones with the
smallest number of poles and zeros on the physical strip. One can
verify directly that $G^{[ab]}(\th)$ introduces no poles onto the physical
strip, so as we have already mentioned the supersymmetric part of the
S-matrix does not introduce any additional bound-state poles.

We must now verify that these S-matrix elements satisfy the bootstrap
equations for the fusing angles in \fao. Schematically these
equations are of the form
$$
S_{\rm P}^{[dc]}(\th)\sim S_{\rm P}^{[da]}(\th-i\bar u_{a\bar c}^{\bar b})
S_{\rm P}^{[db]}(\th+i\bar u_{b\bar c}^{\bar a}).
\efr
Here, $\bar a$ denotes the charge conjugate of $a$, although in the
present section we are assuming $\bar a=a$.
More explicitly a given S-matrix
element can be obtained in two different ways, for example for the
fusing $c=a+b$:
$$ \eqalign{
&S^{[dc]}_{\phi \phi\rightarrow\phi\phi}
(\th)\cr &= S^{[da]}_{\phi \phi\rightarrow\phi\phi}
(\th - i\bar{u}_{a\bar c}^{\bar b})
S^{[db]}_{\phi \phi\rightarrow\phi\phi}(\th + i\bar{u}_{b\bar c}^{\bar
a})+{(f_{\psi
\psi})_{ab}^c\over (f_{\phi \phi})_{ab}^c}
S^{[da]}_{\phi\phi\rightarrow\phi\phi}(\th -
i\bar{u}_{a\bar c}^{\bar b})S^{[db]}_{\psi\phi\rightarrow\psi\phi}
(\th + i\bar{u}_{b\bar c}^{\bar a}) \cr  
&= S^{[da]}_{\phi\psi\rightarrow\psi\phi}(\th - i\bar{u}_{a\bar
c}^{\bar b})S^{[db]}_{\phi\psi\rightarrow\psi\phi}
(\th + i\bar{u}_{b\bar c}^{\bar a})
+{(f_{\phi\phi})_{ab}^c\over (f_{\psi \psi})_{ab}^c} 
S^{[da]}_{\phi\psi\rightarrow\phi\psi}(\th -
i\bar{u}_{a\bar c}^{\bar b})S^{[db]}_{\psi\psi\rightarrow\phi\phi}
(\th + i\bar{u}_{b\bar c}^{\bar a}).\cr}  
\efr
By writting down all 16 of the possible fusings one can directly verify that
the bootstrap equations are consistent with supersymmetry and one
finds the following fusing relation for the unitarizing/crossing
factors:
$$
G^{[d,a+b]}(\th)=\xi^d_{ab}(\th)
G^{[da]}\left(\th-{i\pi b\over H}\right)G^{[db]}\left(\th+{i\pi
a\over H}\right).
\nfr{con}
with
$$
\xi_{ab}^d(\theta)={\sin({\theta\over i})\left[\sin({\theta\over i}+
{(a-b)\pi\over H})+\sin({d\pi\over H})\right]\over \sin({\theta\over i}+
{a\pi\over H})\sin({\theta\over i}-{b\pi\over H})}.
\nfr{myf}
It is a tedious but straightforward excerise to verify that the
solution in \niania\ satisfies the fusion equation \con.

The bootstrap equations arising from the second fusion angle in
\fao\ can be checked in a similar way.
Following the same strategy as before, we find that each one of
the new sixteen bootstrap equations is satisfied if (we take $b\geq a$)
$$ 
G^{[d,b-a]}(\th)=\xi_{-a,b}^{-d}(\th)G^{[db]}\left(\th-{i\pi a\over H}\right)
G^{[da]}\left(\th+{i\pi(H-b)\over H}\right).
\nfr{marco}
This follows from \con\ using crossing symmetry and unitarity.

To summarize the results of this section, we have demonstrated that
the S-matrix describing the scattering of particle super-multiplets 
satisfies bootstrap equations which are schematically of the form
$$\eqalign{
{\rm (i)}\qquad\qquad S_{\rm P}^{[d,a+b]}(\th)&\sim S_{\rm P}^{[da]}\left
(\th-{i\pi b\over H}\right)
S_{\rm P}^{[db]}\left(\th+{i\pi a\over H}\right)\cr
{\rm (ii)}\qquad\qquad S_{\rm P}^{[d,b-a]}(\th)&\sim S_{\rm P}^{[da]}
\left(\th-{i\pi a\over H}\right)
S_{\rm P}^{[db]}\left(\th+{i\pi(H-b)\over H}\right),\cr}
\nfr{tboot}
where in the second equation $b\geq a$.

\section{Fusing between kinks}
 
It is natural at this stage to ask what happens if the bosonic
theory contains states whose mass does not respect the crucial
condition \key\ at the fusing points? These states cannot
correspond to ordinary particle super-multiplets if the fusing rules
of the bosonic theory are to be respected. We shall find that in these
situations the states must be kink super-multiplets of the type
discussed in section 2.2.

The construction of a minimal supersymmetric S-matrix for the scattering of
kinks, $\{K_{0\half},K_{1\half},
K_{\half 0},K_{\half 1} \}$ was described in section
2.2. The S-matrix described there does not have any
poles on the physical strip and so as it stands there are
no kink-kink bound-states. However, the bosonic part of the S-matrix
will introduce poles and so we must consider the possibility of
kink-kink bound-states. Notice that all the possible 
two-kink states carry zero topological
charge and hence it natural to suppose that the bound-states of kinks
correspond to particles. In the following, we shall indicate how such an
indentification is consistent with the action of supersymmetry and
furthermore the scattering of the bound-states, as deduced from 
the bootstrap equations, is identical to that for the particle
super-multiplets written down in section 2.1.

Suppose that the fusing rules imply that two kinks form a bound-state
at a rapidity difference $i\xi$, with 
$0<\xi<\pi$. The bound-state is identified with
a particle super-multiplet $(\phi,\psi)$ of mass
$$
2m\cos(\xi/2).
\nfr{MBS}
It is important that the fusion is consistent for
{\it any\/} value of $\xi$, subject to the constraint that the pole is on
the physical strip. Thus it is possible that a series of particle
multiplets can appear as bound-state of two kinks. Our full solutions
of the bootstrap equations will have this property. 

The coupling constants for these processes are defined via:
$$
\ket{ K_{ab}(\th+i\xi/2)K_{bc}(\th-i\xi/2)}=
f_{abc}^\phi\ket{\phi(\th)}+f_{abc}^\psi\ket{\psi(\th)}.
\efr
The non-zero coupling constants are
$$\eqalign{
f_{0\half0}^\phi=f_{1\half1}^\phi=2^{(\pi-2\xi)/4\pi}
f_{\half0\half}^\phi=2^{(\pi-2\xi)/4\pi}f_{\half1\half}^\phi=
\sqrt{K(i\xi)2^{(\pi-\xi)/2\pi} 
\cos\left({\xi-\pi\over4}\right)}\cr
f_{1\half0}^\psi=-f_{0\half1}^\psi=2^{(\pi-2\xi)/4\pi}i
f_{\half0\half}^\psi=-2^{(\pi-2\xi)/4\pi}if_{\half1\half}^\psi=
\sqrt{K(i\xi)2^{(\pi-\xi)/2\pi} 
\cos\left({\xi+\pi\over4}\right)}.\cr}
\nfr{KCC}
One can easily verify that the relation between the coupling constants
\KCC\ and the S-matrix elements in \CCSM\ is satisfied.

The coupling constants in \KCC\ are consistent with the action of
supersymmetry and fermionic parity. For example
$$\eqalign{
{\cal Q}\ket{\phi(\th)}&={\sqrt m\over f_{0\half0}^\phi}(Q_1+Q_2)
\ket{ K_{0\half}(\th+i\xi/2)K_{\half 0}(\th-i\xi/2)}\cr
&={\sqrt me^{\th/2}\over f_{0\half0}^\phi}\left(-ie^{i\xi/4}
+e^{-i\xi/4}\right)
\ket{ K_{1\half}(\th+i\xi/2)K_{\half 0}(\th-i\xi/2)}\cr
&={e^{-i\pi/4}e^{\th/2}\sqrt{2m\cos(\xi/2)}\over 
f_{1\half0}^\psi}\ket{ K_{1\half}(\th+i\xi/2)K_{\half
0}(\th-i\xi/2)}\cr
&=e^{-i\pi/4}e^{\th/2}\sqrt{2m\cos(\xi/2)}\ket{\psi(\th)},\cr}
\efr
which is exactly the required transformation for a particle of mass \MBS.

As we alluded to earlier, the bound-state structure is consistent for any
value of $\xi$. In the theories that we shall construct in following sections
each of the particle multiplets with masses parameterized as in \gsp\ will 
appear as a kink-kink bound-state for different values of $\xi$. By comparing
\MBS\ with \gsp\ we see that for the $a^{\rm th}$ particle multiplet we have
$\xi=\pi-2\pi a/H$. The scattering of these states can then be determined by
the bootstrap equations. Notice that since a given state, say $\phi_a$,
can be obtained in a number of different ways, for example as a bound-state
of $K_{0\half}$ with $K_{\half0}$ or $K_{1\half}$ with $K_{\half1}$. Hence the
scattering amplitudes of $\phi_a$ which follow from the bootstrap equations
can be obtained in different ways. This leads to a highly 
non-trivial check of the whole construction. Not only do the coupling \KCC\
pass this check but the resulting amplitudes for the particles are
exactly those that we wrote down in section 3.1

To illustrate the construction of particle scattering amplitudes from the
kink amplitudes via the bootstrap equations consider the process
$\phi_a\phi_b\rightarrow\phi_b\phi_a$ computed by using the fusion
$$
\ket{ K_{0\half}(\th+i\xi/2)K_{\half0}(\th-i\xi/2)}=\sqrt{K(i\xi)
2^{(\pi-\xi)/2\pi}\cos\left({\xi-\pi\over4}\right)}\ket{\phi(\th)}.
\efr
with $\xi=\pi-2\pi a/H$ for $\phi_a$ and $\xi=\pi-2\pi b/H$ for $\phi_b$, 
respectively. The bootstrap equation is illustrated in figure 1.

\sjump
\vbox{
\centerline{
\psfig{figure=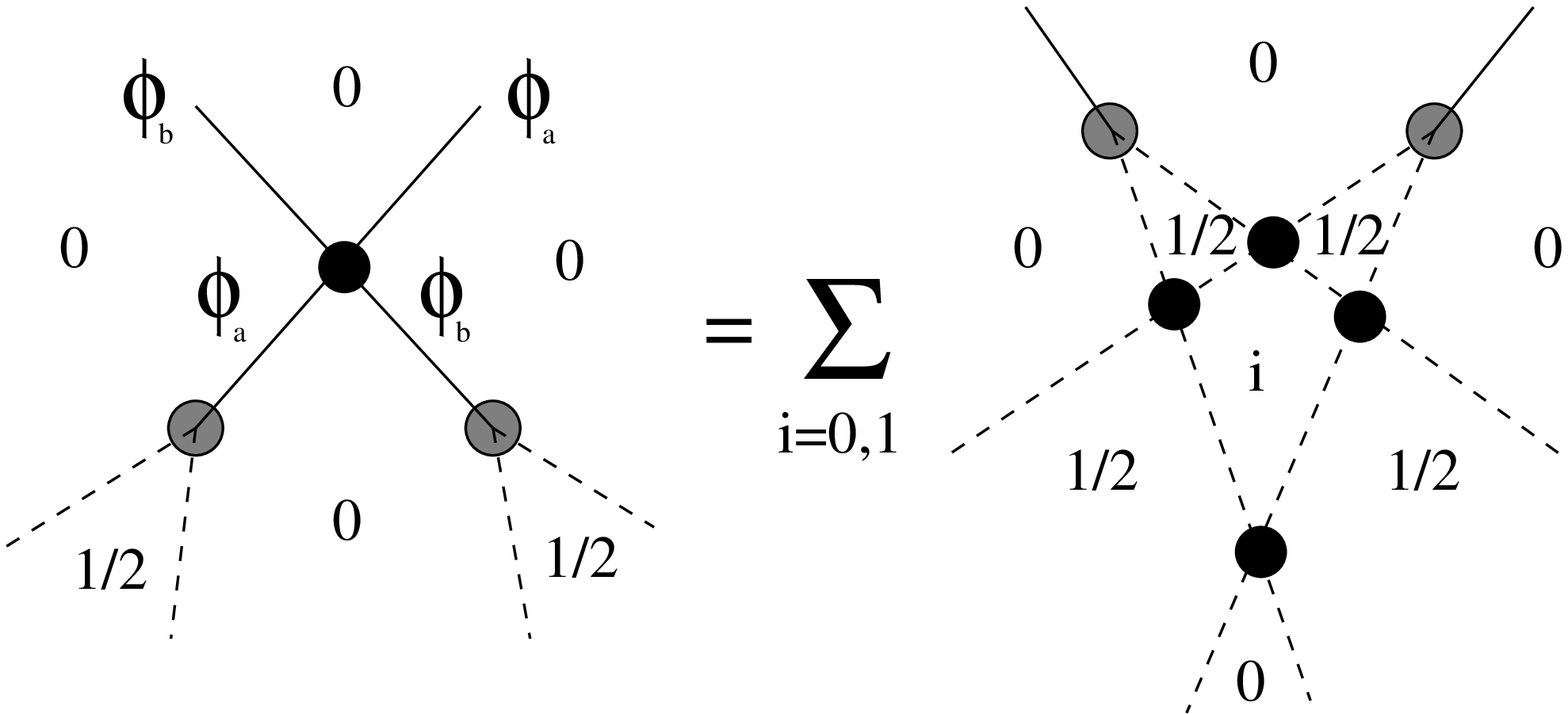,height=5cm}}
\bjump
\centerline{Figure 1. Particle S-matrix element in terms of kink S-matrices}
}

\sjump
Hence the scattering amplitude for $\phi_a$ with $\phi_b$ is equal to
$$\eqalign{
&S\left(\left.\matrix{0&\half\cr \half&0\cr}\right\vert\th+ix\right) 
S\left(\left.\matrix{\half&0\cr 0&\half\cr}\right\vert\th-iy\right) 
S\left(\left.\matrix{0&\half\cr \half&0\cr}\right\vert\th-ix\right) 
S\left(\left.\matrix{\half&0\cr 0&\half\cr}\right\vert\th+iy\right) \cr
\qquad&+
S\left(\left.\matrix{1&\half\cr \half&0\cr}\right\vert\th+ix\right) 
S\left(\left.\matrix{\half&1\cr 0&\half\cr}\right\vert\th-iy\right) 
S\left(\left.\matrix{0&\half\cr \half&1\cr}\right\vert\th-ix\right) 
S\left(\left.\matrix{\half&0\cr 1&\half\cr}\right\vert\th+iy\right) \cr
&=K(\th+ix)K(\th-iy)K(\th-ix)K(\th+iy)
\left({1\over2}\sin(\th/i)+\cos(y/2)\cos(x/2)\right),\cr}
\efr
where $x=(a-b)\pi/H$ and $y=\pi-(a+b)\pi/H$. This is precisely equal
to $S_{\phi\phi\rightarrow\phi\phi}^{[ab]}(\th)$ in \rgo\ by virtue of
the following the identity between the unitarizing/crossing factors: 
$$
G^{[ab]}(\th)=
{\sin(\th/i)\over2}K(\th+ix)K(\th-iy)K(\th-ix)K(\th+iy)
\efr

The S-matrix elements for the scattering of kinks with particles also
follows by applying the bootstrap equations. For example consider the
process $\phi_a(\th_1)+K_{0\half}(\th_2)\rightarrow
K_{0\half}(\th_2)+\phi_a(\th_1)$. This S-matrix element can be deduced from
the bootstrap equation illustrated in figure 2 and is equal to
$$
{f_{\half0\half}^{\phi_a}\over f_{0\half0}^{\phi_a}}
S\left(\left.\matrix{0&\half\cr \half&0\cr}\right\vert\th+ix\right) 
S\left(\left.\matrix{\half&0\cr 0&\half\cr}\right\vert\th-ix\right)
+{f_{\half1\half}^{\phi_a}\over f_{0\half0}^{\phi_a}}
S\left(\left.\matrix{0&\half\cr \half&1\cr}\right\vert\th+ix\right) 
S\left(\left.\matrix{\half&1\cr 0&\half\cr}\right\vert\th-ix\right),
\efr
where $x=\pi/2-\pi a/H$ and the coupling constants are given in \KCC\
with $\xi=\pi-2\pi a/H$.

\sjump
\vbox{
\centerline{
\psfig{figure=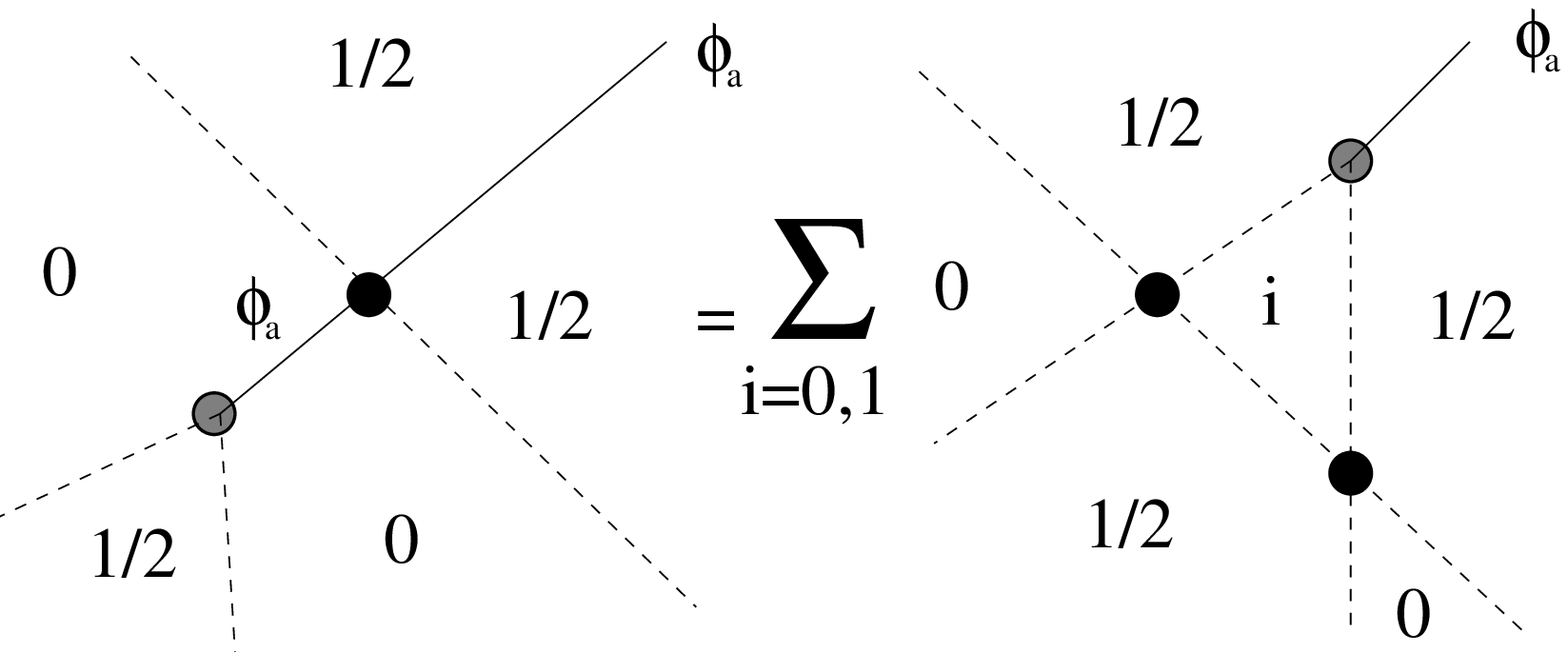,height=4cm}}
\bjump
\centerline{Figure 2. Bootstrap gives particle-kink scattering}
}

\sjump
\section{Fusing between particles and kinks}

Since two kinks can fuse at a rapidity difference $i\xi$ to form a particle, 
it is also possible for a particle and kink to fuse to form a kink at a
rapidity difference $i\pi-i\xi/2$. This allows to introduce the
particle-kink coupling constants:
The coupling constants for these processes are defined via:
$$
\ket{\varphi(\th+i\xi/2)K_{ab}(\th-i\pi+i\xi)}=
f_{\varphi ab}^c\ket{K_{cb}(\th)},
\efr
where $\varphi\in(\phi,\psi)$. 
The non-zero coupling constants are
$$\eqalign{
&f_{\phi0\half}^0=f_{\phi1\half}^1=2^{(\pi-2\xi)/4\pi}
f_{\phi\half0}^\half=2^{(\pi-2\xi)/4\pi}f_{\phi\half1}^\half=
\sqrt{K(i\xi)2^{(\pi-\xi)/2\pi} \cos\left({\xi-\pi\over4}\right)}\cr
&f_{\psi1\half}^0=-f_{\psi0\half}^1=2^{(\pi-2\xi)/4\pi}i
f_{\psi\half1}^\half=-2^{(\pi-2\xi)/4\pi}if_{\psi\half0}^\half=
\sqrt{K(i\xi)2^{(\pi-\xi)/2\pi}\cos\left({\xi+\pi\over4}\right)}.\cr}
\nfr{KPCC}
Notice that the coupling constant are simply the the crossed versions of
those in \KCC. It is furthermore possible to verify by explicity constructing
all the particle-kink S-matrix elements as illustrated at the end of the
last section, that the relation between the coupling constants
\KPCC\ and the S-matrix elements in \CCSM\ is satisfied.
It is also possible to show that the 
bound-states have the correct transformation
properties under supersymmetry and fermionic-parity for {\it any\/} value of 
$\xi$. 

The fusions in \KPCC\ imply a set of bootstrap equations. It is at this stage
that the bootstrap equations close in on themselves because if a kink can be
formed as a bound-state of a particle and a kink and a particle can be formed
as a bound-state of two kinks, then it follows that a kink can be formed as
a bound-state of three kinks. Therefore the resulting bootstrap equations
are a highly non-trivial set of consistency conditions on the kink S-matrix
elements. To illustrate this, consider the bootstrap equation depicted in
figure 3. 

\sjump
\vbox{
\centerline{
\psfig{figure=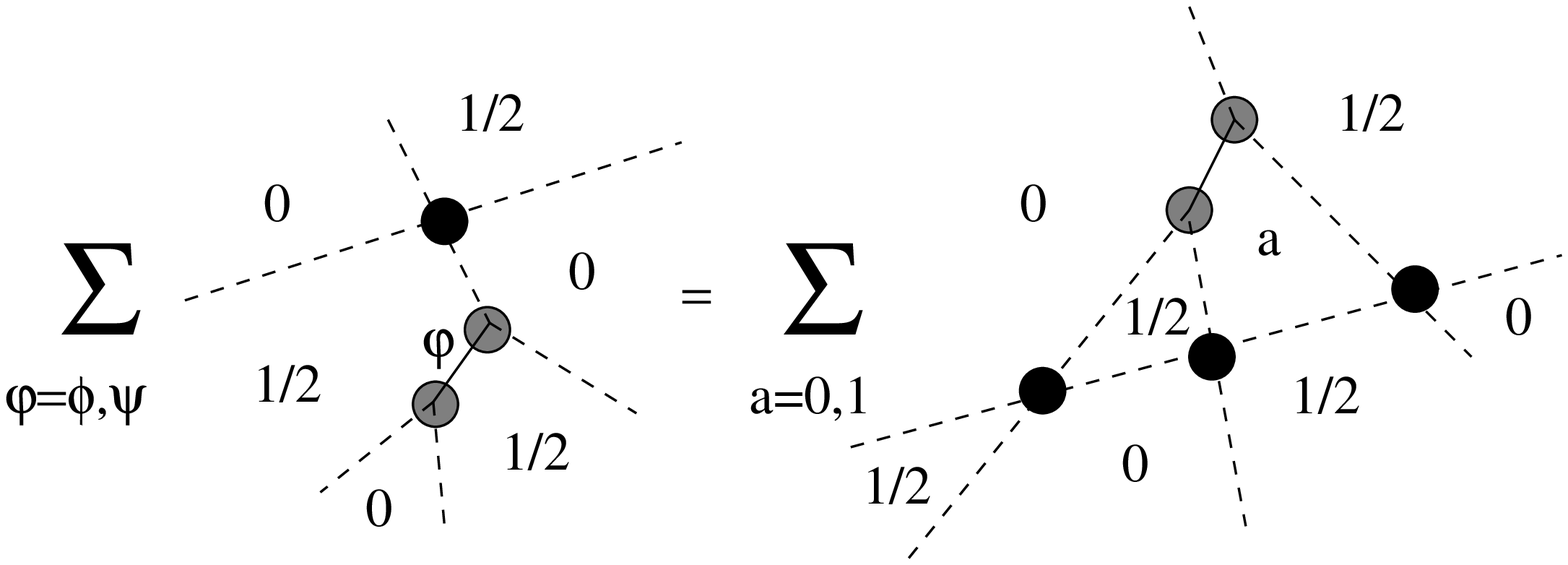,height=4.5cm}}
\bjump
\centerline{Figure 3. Non-trivial bootstrap equation for kinks}
}

\sjump
This requires the following relation between the kink S-matrix
elements:
$$\eqalign{
S\left(\left.\matrix{0&\half\cr \half&0\cr}\right\vert\th\right)&=
\alpha(\xi)
S\left(\left.\matrix{0&\half\cr \half&0\cr}\right\vert\th_1\right) 
S\left(\left.\matrix{\half&0\cr 0&\half\cr}\right\vert\th_2\right) 
S\left(\left.\matrix{0&\half\cr \half&0\cr}\right\vert\th_3\right)\cr
&\qquad\qquad +\beta(\xi)S\left(\left.\matrix{0&\half\cr
\half&0\cr}\right\vert\th_1
\right) 
S\left(\left.\matrix{\half&1\cr 0&\half\cr}\right\vert\th_2\right) 
S\left(\left.\matrix{1&\half\cr \half&0\cr}\right\vert\th_3\right)\cr}
\nfr{TED}
where $\th_1=\th-i\xi$, $\th_2=\th$ and
$\th_3=\th+i\pi-i\xi$. The quantities $\alpha(\xi)$ and 
$\beta(\xi)$ are the following functions of the coupling constants:
$$\eqalign{
\alpha(\xi)&={f_{0\half0}^\phi f_{\phi(0\half)}^0\over
f_{\half0\half}^\phi f_{\phi\half0}^\half+
f_{\half0\half}^\psi f_{\psi\half0}^\half}={\sqrt2\cos((\xi-\pi)/4)\over
\cos(\xi/4)}2^{-\xi/\pi}\cr
\beta(\xi)&={f_{0\half0}^\psi f_{\psi1\half}^0\over
f_{\half0\half}^\phi f_{\phi\half0}^\half+
f_{\half0\half}^\psi f_{\psi\half0}^\half}={\sqrt2\cos((\xi+\pi)/4)\over
\cos(\xi/4)}2^{-\xi/\pi}.\cr}
\efr
It is a tedious exercise to verify that the S-matrix elements of the
kinks do indeed satisfy \TED\ and similar equations that result from
considering other possible fusions. The calculation is greatly
aided by noting that
$$
K(\th-i\xi)K(\th+i\pi-i\xi)={1\over\cos((\th-i\xi)/2i)},
\efr
which follows from using the unitarity and crossing properties of
$K(\th)$. It is important to realize that these bootstrap
equations are satisfied for any value of the 
fusing angle $\xi$. 

\chapter{Supersymmetric S-matrices associated to Lie algebras}

The simplest series of bosonic factorizable S-matrices are 
purely elastic---the particles carry no internal quantum 
numbers---and are related to Lie algebra data. In
the case of a simply-laced algebra there is a so-called minimal
S-matrix which does not depend on any coupling constant. By
introducing specific CDD factors depending on a single constant 
these S-matrices describe the scattering of particle states in
affine Toda field theories. There is an elegant group theoretical way
to write these S-matrices due to Dorey [\Ref{D1}]
For the non-simply-laced algebras, the
minimal S-matrix itself depends on a coupling constant and they
directly describe affine Toda field theories [\Ref{CDS}]. 
Below, we consider each
of the classical affine Lie algebras in turn and how a supersymmetric
S-matrix can be associated to each of them. For these S-matrices the
factorization of the S-matrix implied by \FAC\ is somewhat redundant
since the bosonic factor can be absorbed as a CDD factor into the
definition of the supersymmetric factor; however, the factorized form
makes the fusion structure of the theories manifest.

Before doing so, it is useful to introduce the following
standard notation used for the bosonic theories [\Ref{BCDS}]:
$$
\eqalign{
(x)&={\sin \left( {\th\over 2i} +{\pi x\over 2h} \right)\over \sin
\left( {\th\over 2i} -{\pi x\over 2h} \right)},\quad
\{x\}= (x-1)(x+1),\cr
\{x \} _v &={(x-vB-1)(x+vB+1)\over (x+vB+B-1)(x-vB-B+1)}, \qquad 
[x] _v =\{ x \} _v \{ h-x \} _v.\cr}
\efr

\section{The case $a_{n-1}^{(1)}$}

The purely elastic S-matrix associated to this algebra
involves a set of $n-1$ particles with masses [\Ref{AFZ}]
$$
m_a=m\sin\left({\pi a\over n}\right),\ a=1,2,\ldots,n-1.
\efr
The fusion angles are
$$
u_{ab}^{a+b}=
{(a+b)\pi\over n}\quad(a+b<n),\qquad
u_{ab}^{a+b-n}={(2n-a-b)\pi\over n}\quad(a+b>n),
\nfr{anf}
and the minimal S-matrix elements are given by
$$
\tilde S^{[ab]}(\th)=\prod_{j= |a-b|+1\atop{\rm step}\
2}^{|a+b|-1}\{j\}.
\efr
The affine Toda S-matrix is given by replacing $\{j\}$ with
$\{j\}_0$ where the coupling constant $0<B<2$.

From both the minimal and affine Toda 
S-matrix, a supersymmetric S-matrix with $n-1$
particle super-multiplets can be built by appending the supersymmetric factors
$S_{\rm P}^{[ab]}(\th)$. In order to prove that this is consistent 
we must show
that the supersymmetric part of the S-matrix satisfies the bootstrap
equations which result from the fusing angles in \anf. Before we can
use the results in section 3.1 we must first take account of the fact
that in this theory the particles are not self-conjugate, as assumed
there, rather $\bar a=n-a$. Hence the crossing symmetry
relation in \uac\ should be replaced by 
$$
G^{[ab]}(\th)=G^{[n-a,b]}(i\pi-\th).
\nfr{ccg}
It is easy to verify that the general solution \niania\ does satisfy
the above relation with $H=n$. 

Consider the bootstrap equations that arise from the fusing rules
\anf. If $a+b<n$, we have already shown that $S_{\rm P}(\th)$
satisfies the corresponding bootstrap equation, namely (i) of \tboot. 
When $a+b\geq n$, the corresponding bootstrap equation is (schematically)
$$
S_{\rm P}^{[d,a+b-n]}(\th)\sim 
S_{\rm P}^{[da]}\left(\th-{i\pi(n-b)\over n}\right)
S_{\rm P}^{[db]}\left(\th+{i\pi(n-a)\over n}\right).
\efr
Now using the fact that 
$$
S_{\rm P}^{[ab]}(\th)\equiv S_{\rm P}^{[n-a,b]}(\th), 
\efr
which follows from \ccg\ and \rgo, we can express the above as
$$
S_{\rm P}^{[d,2n-a-b]}(\th)\sim S_{\rm
P}^{[d,n-a]}\left(\th-{i\pi(n-b)\over n}\right)
S_{\rm P}^{[d,n-b]}\left(\th+{i\pi(n-a)\over n}\right).
\efr
But this is precisely the bootstrap equation (i) of \tboot\ for 
$(n-a)+(n-b)\rightarrow(2n-a-b)$ which has already been verified.
Hence the
supersymmetric part of the S-matrix respects all the fusing of the
bosonic theory and so the combined S-matrix $\tilde
S^{[ab]}(\th)S_{\rm P}^{[ab]}(\th)$ is a consistent supersymmetric S-matrix. 

\section{The case $d_{n}^{(1)}$}

The bosonic S-matrix related to this algebra describes a theory with $n$
particles with masses [\Ref{CDS}]
$$
\eqalign{
m_{s}&=m_{s'}=m,\cr
m_a&=2m \sin \left( {a \pi\over 2(n-1)}\right) \quad
a=1,2,\ldots,n-2.\cr}
\nfr{mdn}
Here we have labelled the particles $n-1$ and $n$ as $s$ and $s'$,
because they are associated to the spinorial spots of the $d_n$
Dynkin diagram. We will refer to these two particles as the spinor and
anti-spinor. For $n$ even all the particles are self-conjugate,
whereas for $n$ odd $\bar s=s'$, and vice-versa.

The fusion angles between non-spinor particles are
$$
\eqalign{
&u_{ab}^{a+b}={(a+b)\pi\over2(n-1)}\quad(a+b\leq n-2),\qquad
u_{ab}^{|b-a|}={(2(n-1)-|b-a|)\pi\over2(n-1)},\cr
&u_{ab}^{2(n-1)-a-b}={(a+b)\pi\over2(n-1)}\quad(a+b\geq n).\cr}
\nfr{dfo}
Those involving the spinor particles are 
$$
\eqalign{
&u_{ss}^a=u_{s's'}^a={(n-1-a)\pi\over n-1},\ \ u_{sa}^{s}=u_{s'a}^{s'}=
{(n-1+a)\pi\over2(n-1)},\cr
&u_{ss'}^a={(n-1-a)\pi\over n-1},\ \ u_{s'a}^{s}=u_{sa}^{s'}=
{(n-1+a)\pi\over2(n-1)}.\cr}
\nfr{dft}
In the first line of the above equation $a$ is restricted to be even
if $n$ is even and $a$ is odd if $n$ is odd, and vice-versa in the
second line.

The minimal S-matrix elements involving the non-spinor particles are
$$
\tilde S^{[ab]}(\th)=\prod_{j= |a-b|+1\atop{\rm step}\ 2}^{|a+b|-1}
\{j\}\{2n-2-j\}.
\efr
The elements for the scattering of a spinor particle with a non-spinor
particle are
$$
\tilde S^{[sa]}(\th)=\tilde S^{[s'a]}(\th)=\prod_{j= 0\atop{\rm step}\
2}^{2a-2}\{n-a+j\}.
\efr
Finally, the S-matrix elements involving only spinor particles are
$$
\eqalign{
\tilde S^{[ss]}(\th)=\tilde S^{[s's']}(\th)&=
\prod_{j= 1\atop{\rm step}\ 4}^{2n-3}\{j\},\ \  
\tilde S^{[ss']}(\th)=\prod_{j= 3\atop{\rm step}\ 4}^{2n-5}\{j\},\ \ n\ {\rm
even}\cr
\tilde S^{[ss]}(\th)=\tilde S^{[s's']}(\th)&=
\prod_{j= 1\atop{\rm step}\ 4}^{2n-5}\{j\},\ \  
\tilde S^{[ss']}(\th)=\prod_{j= 3\atop{\rm step}\ 4}^{2n-3}\{j\},\ \ n\ 
{\rm odd}.\cr}
\efr
As in the $a_{n-1}^{(1)}$ example the S-matrix of the affine Toda
field theory is obtained by replacing $\{j\}$ with $\{j\}_0$ where the 
coupling constant $0<B<2$.

It is immediately apparent from the masses \mdn\ that in this case the
coupling constant $H$ is fixed to be $2(n-1)$, the Coxeter number of
$d_n$. In addition, we see that in the 
supersymmetric construction the spinor particles must be kink
states, whereas the other states will be particles. So we introduce
two sets of kinks: $\{ K_{0, \half}^{s},
K_{1, \half}^{s}, K_{ \half , 0}^{s}, K_{ \half
1}^{s}\}$ and $\{ K_{0, \half}^{s'},K_{1, \half}^{s'},
K_{ \half , 0}^{s'}, K_{ \half 1}^{s'} \}$.
The scattering of the kink degrees-of-freedom is governed by the kink
S-matrix described in section 2.2.

We must now verify that the supersymmetric part of the S-matrix
respects the bootstrap equations which follow from the fusing angles
in \dfo\ and \dft. The first two cases in \dfo\
follow immediately from \tboot. The
bootstrap equation corresponding to the third fusing rule in
\dfo\ is
$$
S_{\rm P}^{[d,2(n-1)-a-b]}(\th)\sim
S_{\rm P}^{[da]}\left(\th-{ib\pi\over2(n-1)}\right)S_{\rm P}^{[db]}\left(
\th+{ia\pi\over2(n-1)}\right).
\nfr{num}
Although the non-spinor particle label runs from $1$ to $n-2$ we can
artificially extend the label to run to $2(n-1)$ by identifying
$a$, with $1\leq a\leq n-2$ with $2(n-1)-a$. This identification is
consistent with S-matrix elements defined on the extended labels since
$$
S_{\rm P}^{[ab]}(\th)=S_{\rm P}^{[a,2(n-1)-b]}(\th)=
S_{\rm P}^{[2(n-1)-a,b]}(\th)=S_{\rm P}^{[2(n-1)-a,2(n-1)-b]}(\th).
\nfr{blob}
The bootstrap equations corresponding to the first fusing rule in
\dfo\ are also satisfied by the S-matrices defined on the extended set
of labels:
$$
S_{\rm P}^{[d,a+b]}(\th)\sim 
S_{\rm P}^{[da]}\left(\th-{ib\pi\over2(n-1)}\right)
S_{\rm P}^{[db]}\left(\th+{ia\pi\over2(n-1)}\right).
\efr
Now if $a+b\geq n$ then by using \blob\ we see that \num\ is
satisfied.

Now we move on to consider scattering amplitudes containing
kinks. Although we have introduced two sets of kinks associated to the
spinor particles the situation is quite simple because the
supersymmetric part of the S-matrix is blind to the labels $s$ and $s'$.
Bound-states corresponding to particles exist in the $ss$, $ss'$
or $s's'$ channels with fusion angles given in \dft. 
The associated bootstrap equations are satisfied by virtue of the
discussion in sections 3.2 and 3.3. The important point to remember is that the
supersymmetric part of the S-matrix satsifies the required bootstrap equation
for {it any\/} value of the kink-kink fusing angle denoted $\xi$
in section 3.2.

\section{The pair $c_n^{(1)}$, $d_{n+1}^{(2)} $}

This is the first example involving non-simply-laced algebras [\Ref{CDS}]. In
these cases each of the S-matrices is associated to a pair of
algebras. In addition, the S-matrices and the mass spectrum depend
on a coupling constant, and there is no analogue of the minimal
coupling constant independent S-matrix that exists in the simply-laced
cases.

In the present case the mass spectrum is
$$ 
m_a=2m\sin\left({a\pi\over H}\right),\quad p=1,2,\ldots,n,
\nfr{cn1}
where the coupling constant $2n\leq H\leq 2n+2$. The particles are all
self-conjugate and the fusion angles are
$$
u_{ab}^{a+b}={(a+b)\pi\over H}\quad(a+b\leq n),\qquad
u_{ab}^{|b-a|}={(H-|b-a|)\pi\over H}.
\efr
The S-matrix elements are
$$ 
\tilde S^{[ab]}(\th)=\prod _{j=|a-b|+1\atop{\rm step}\ 2}^{
a+b-1}[j] _0.
\efr

In this case the mass spectrum is exactly that of \gsp\ and so we can
build a supersymmetric S-matrix by simply appending the particle
supersymmetric S-matrix of section 2.1 to the bosonic S-matrix, giving
$\tilde S^{[ab]}(\th)S^{[ab]}_{\rm P}(\th)$.

\section{The pair $b^{(1)}_n$, $a^{(2)}_{2n-1}$}

In this case the spectrum contains $n$ particles of mass [\Ref{CDS}]
$$
m_a=2m\sin\left({a\pi\over H}\right),\qquad a=1,2,\ldots,n-1,\qquad m_n=m.
\efr
As in the previous example, the
quantity $H$ can float in the region $2n-1\leq H\leq 2n$. The
particles are all self-conjugate and the fusion angles are
$$
\eqalign{
u_{ab}^{a+b}&={(a+b)\pi\over H}\quad(a+b<n),\qquad
u_{ab}^{|b-a|}={(H-|b-a|)\pi\over H}\cr
u_{nn}^a&={(H-2a)\pi\over H},\ u_{na}^n={(H/2+a)\pi\over H}.\cr} 
\efr
The S-matrix elements are
$$
\eqalign{
\tilde S^{[ab]}(\th)&=\prod _{j=|a-b|+1\atop{\rm step}\ 2}^{
a+b-1}[j]_0\cr
\tilde S^{[an]}(\th)&=\prod _{j=1\atop{\rm step}\ 2}^{
2a-1}\{H/2-a+j\}_0\cr
\tilde S^{[nn]}(\th)&=\prod _{j=1-n\atop{\rm step}\ 2}^{
n-1}\{H/2-j\}_{-1/4}.\cr}
\efr

From the mass formula, it is clear that the particle $n$,
corresponding to the spinor spot of the $b_n$ Dynkin diagram, should
be associated to a kink multiplet in the supersymmetric theory. The
remaining particles become ordinary particle supermultiplets in the 
supersymmetric theory. The bootstrap equations involving just these
particles follows from the results of section 3.1. The bootstrap
equations involving the kinks are verified in the same way as in the
$d_n^{(1)}$ theory.

\chapter{Other theories}

In this section we briefly consider the S-matrices of other well-known
supersymmetric integrable theories.

\section{The O($2n$) supersymmetric sigma model}

The lagrangian density of the supersymmetric O($2n$) model is [\Ref{SW}]
$$
{\cal L}={1\over2g}\left[(\partial_\mu
n_a)^2+i\bar\psi_a\slashchar\partial\psi_a+\frac{1}{4}\left(
\bar\psi_a\psi_a\right)^2\right],
\nfr{lag}
where $n_a$ and $\psi_a$ are a $2n$-component real scalar field and 
Majorana fermion, respectively, satisfying the constraints
$n\cdot n=1$ and $n\cdot\psi=0$. 
The theory \lag\ has a global O($2n$)
symmetry and a global $N=1$ supersymmetry. Notice that the bosonic
part of the theory is just the O($2n$) sigma model, the fermionic
part is the O($2n$) Gross-Neveu model, and the coupling between the bosons and
fermions is due solely to the constraint.

The $S$-matrix conjectured by Shankar and Witten [\Ref{SW}] 
for the scattering of the fundamental
states, which transform in the vector representation of O($N$) and
form a particle super-multiplet of supersymmetry,
has the factored form of \FAC. In this case the bosonic factor $\tilde
S(\th)$
is the S-matrix of vector state of the O($2n$) Gross-Neveu model
[\Ref{ZZ},\Ref{SMGN}] and the supersymmetric part is precisely the
S-matrix block of a particle super-multiplet written down explicitly 
in section 3.1. 

The outstanding question is 
what is the full S-matrix of the theory generated by
the bootstrap programme. Since the
supersymmetric part introduces no additional poles onto the physical
strip it is natural to suppose that the full S-matrix enjoys the same
fusing angles as the O($2n$) Gross-Neveu model. This theory has a
spectrum and fusing date 
which is exactly that of the $d_{n}^{(1)}$ model discussed in section
4.2, where the particle of mass $m_1$ transforms in the vector
representation and the heavier particles $m_a$, for $a=2,\ldots,n-2$,
transform in reducible representations. The two `spinor' particles
transform in the spinor and anti-spinor representations of O($2n$).
In the supersymmetric theory the particles with masses $m_a$, for
$a=1,\ldots,n-2$, become particle super-multiplets, while the spinor
particles become kink super-multiplets. 

The fact that the resulting S-matrix satisfies the bootstrap
equations, follows from the discussion in section 4.2. 
The S-matrix conjectured by Shankar and Witten for the scattering of
the vector states has been subjected to a highly non-trivial test
using the Thermodynamic Bethe Ansatz (TBA) [\Ref{EH1}]. Not only is the
conjectured S-matrix completely consistent with the TBA calculation
but as a by-product an exact value for the mass-gap of the model is
found (see [\Ref{EH2}] and references therein).
The case $n=2$ deserves a special mention because in this case the
theory is actually the supersymmetric SU(2) principal chiral model. It
has been argued that the spectrum of this theory should include only
the spinor states, i.e. only the two kink super-multiplets which
transform as $(2,1)+(1,2)$ of the SU(2)$\times$SU(2) symmetry. Our
S-matrix is valid in this case since the poles corresponding to the
particle super-multiplets disappear from the physical strip and the
S-matrix of the kink states is consistent by itself. This model is the
first of the SU($n$) series of supersymmetric principal chiral models
whose S-matrices have been written down in [\Ref{EH3}] and subjected
to the highly non-trivial TBA test. 

\section{The supersymmetric sine-Gordon theory}

The exact S-matrix of the supersymmetric sine-Gordon theory
[\Ref{WO},\Ref{SW}] has been
conjectured in [\Ref{A1}]. The spectrum of the model consists of a
soliton and anti-soliton, of mass $m$, which both transform as kink
super-multiplets. Just as in the sine-Gordon theory without
supersymmetry, the soliton and anti-soliton have bound-states called
breathers of mass
$$
m_a=2m\sin\left({a\gamma\over16}\right), 
\efr
where $a=1,2,\ldots,N<8\pi/\gamma$,
and $\gamma$ is a funcion of the coupling constant of the model. 
The breathers transform in particle super-multiplets.  

The S-matrix has the factored form of \FAC\ where the bosonic factor is
the S-matrix of the sine-Gordon theory [\Ref{ZZ}]. 
In order to prove that the
full S-matrix satisfies the bootstrap equations we must consider the
possible fusing rules. Denoting the soliton by $s$ and the
anti-soliton by $s'$ the fusing rules are
$$
u_{ab}^{a+b}={(a+b)\gamma\over16}\quad(a+b\leq N),\qquad
u_{ab}^{|b-a|}=\pi-{|b-a|\gamma\over16},
\nfr{SGP} 
for processes involving the breathers alone and 
$$
u_{ss'}^a=\pi-{a\gamma\over8},\qquad
u_{sa}^s=u_{s'a}^{s'}={\pi\over2}-{a\gamma\over16}.
\nfr{SGK}
for processes involving the solitons. The fusing angles of the
breathers are identical to those in \fao\ and the bootstrap equations 
follow from the results derived in section 3.1 summarized in \tboot.
For the solitons the discussion in section 3.2 implies that 
the bootstrap equations are satisfied.

\chapter{Discussion}

The fact that the S-matrices of many integrable field theories are
known exactly follows from the tractability of the bootstrap
equations. Remarkably, the bootstrap data encoded in the fusing angles
are related to (affine) Lie algebras. It is surprising that when  one
considers integrable theories with $N=1$ supersymmetry that solutions
of the bootstrap can be associated to exactly the same fusing data as
for the bosonic theories. Using this fact we have been able to show
that some well-known supersymmetric integrable theories, namely the
supersymmetric O($2n$) sigma model and the supersymmetric sine-Gordon
model, have exact S-matrices which complete the bootstrap programme. 

In addition we showed how supersymmetric S-matrices could be
associated to the theories with the same fusing data as the affine Toda
field theories. We do not know whether these S-matrices describe any
Lagrangian field theories. For instance they cannot describe the usual
supersymmetric Toda theories, since these are associated to super Lie algebras,
rather than Lie algebras [\Ref{EH5}].

It is important for us to stress that the solutions of the bootstrap
equations for theories with $N=1$ supersymmetry that we have found are not
exhaustive. For instance, recently, the S-matrices of the $N=1$
supersymmetric SU$(n)$ principal chiral models have been found
[\Ref{EH3}]. These S-matrices generalize the kink S-matrices described
here since they are constructed from solutions of the Yang-Baxter
equation for an SU($n$) quantum group. As we have already mentioned,
for $n=2$ this model is the supersymmetric O(4) model.

\bjump
TJH is supported by a PPARC Advanced Fellowship.

\references

\beginref
\Rref{ZZ}{A.B. Zamolodchikov and Al. B. Zamolodchikov, Ann. Phys,
{\bf120} (1979) 253}
\Rref{SW}{R. Shankar and E. Witten, Phys. Rev. {\bf D17} (1978), 2134}
\Rref{WO}{E. Witten and D. Olive, Phys. Lett. {\bf B78} (1978), 97}
\Rref{zam5}{A.B Zamolodchikov, Moscow preprint (1989)}
\Rref{ABL}{C. Ahn, D. Bernard and A. LeClair, Nucl. Phys. {\bf B346} 
(1990) 409}
\Rref{A1}{C. Ahn, Nucl. Phys. {\bf B354} (1991), 57}
\Rref{S1}{K.  Schoutens, Nucl. Phys. {\bf B344} (1990) 665}
\Rref{BCDS}{H. W. Braden,
 E. Corrigan, P.E. Dorey and R. Sasaki, Nucl. Phys. {\bf B388} 689 (1990)}
\Rref{CDS}{E. Corrigan, P.E. Dorey and R. Sasaki, Nucl. Phys. {\bf B408}
579  (1993)\newline
G.W. Delius, M.T. Grisaru and D. Zanon, Nucl. Phys. {\bf B382} (1992) 365}
\Rref{EH1}{J. M. Evans and T. J. Hollowood, Phys. Lett. {\bf B343} (1995) 189}
\Rref{EH2}{J. M. Evans and T. J. Hollowood, Nucl. Phys. {\bf B}
(Proc. Suppl.) {\bf 45A} (1996) 130} 
\Rref{BL}{D. Bernard and A. LeClair, Nucl. Phys. {\bf B340} (1990)
721}
\Rref{EH3}{J.M. Evans and T.J. Hollowood, `{\it Exact scattering in
the SU(n) principal chiral model\/}', {\tt hep-th/9603190}}
\Rref{AFZ}{A.E. Arinshtein, V.A. Fateev and A.B. Zamolodchikov,
Phys. Lett. {\bf B87} (1979) 389}
\Rref{SMGN}{M. Karowski and H.J. Thun, Nucl. Phys. {\bf B190} (1981)
61}
\Rref{D1}{P. Dorey, Nucl. Phys. {\bf B358} (1991) 654} 
\Rref{EH5}{J.M. Evans and T.J. Hollowood, Nucl. Phys. {\bf B352}
(1991) 723}
\endref
\ciao